\documentclass{aa}
\pdfoutput=1
\usepackage{epsfig}
\usepackage{epstopdf}
\usepackage{multirow}
\usepackage[varg]{txfonts}

\usepackage{hyperref}

\def \aap  {A\&A}

\def \apjs  {ApJS}
\def \apjl  {ApJL}

\def \chisq  {\ifmmode  \chi^2   \else  $\chi^2$  \fi}  
\def \spose#1{\hbox  to 0pt{#1\hss}}  
\def \lta{\mathrel{\spose{\lower 3pt\hbox{$\sim$}}\raise  2.0pt\hbox{$<$}}}
\def \gta{\mathrel{\spose{\lower  3pt\hbox{$\sim$}}\raise 2.0pt\hbox{$>$}}}

\def \kms {\ifmmode  \,\rm km\,s^{-1} \else $\,\rm km\,s^{-1}  $ \fi }
\def \kpc {\ifmmode  {\rm~kpc}  \else ${\rm~kpc}$\fi}  
\def \pc {\ifmmode  {\rm~pc}  \else ${\rm~pc}$ \fi  }  
\def \Gyr {\ifmmode  {\rm~Gyr}  \else ${\rm~Gyr}$\fi}
\def \Msun {\ifmmode M_{\odot} \else $M_{\odot}$ \fi} 
\def \Lsun {\ifmmode L_{\odot} \else $L_{\odot}$ \fi} 
\def \Rsun {\ifmmode R_{\odot} \else $R_{\odot}$ \fi} 
\def \Msunpyr {\ifmmode M_{\odot}{\rm~yr}^{-1} \else $M_{\odot}{\rm~yr}^{-1}$ \fi} 
\def \hMsun {\ifmmode h^{-1}\,\rm M_{\odot} \else $h^{-1}\,\rm M_{\odot}$ \fi}

\def \LCDM {\ifmmode \Lambda{\rm CDM} \else $\Lambda{\rm CDM}$ \fi}
\def \sig8 {\ifmmode \sigma_8 \else $\sigma_8$ \fi} 
\def \OmegaM {\ifmmode \Omega_{\rm M} \else $\Omega_{\rm M}$ \fi} 
\def \OmegaL {\ifmmode \Omega_{\rm \Lambda} \else $\Omega_{\rm \Lambda}$\fi} 
\def \Deltavir {\ifmmode \Delta_{\rm vir} \else $\Delta_{\rm vir}$ \fi}
\def \rhocrit {\ifmmode \rho_{\rm crit} \else $\rho_{\rm crit}$ \fi}
\def \rhou {\ifmmode \rho_{\rm u} \else $\rho_{\rm u}$ \fi}
\def \zc {\ifmmode z_{\rm c} \else $z_{\rm c}$ \fi}

\def \rhos {\ifmmode \rho_{\rm s} \else $\rho_{\rm s}$ \fi} 
\def \rs {\ifmmode r_{\rm s} \else $r_{\rm s}$ \fi} 
\def \cvir {\ifmmode c_{\rm vir} \else $c_{\rm vir}$ \fi} 
\def \Rvir {\ifmmode r_{\rm vir} \else $R_{\rm vir}$ \fi}
\def \Vvir {\ifmmode V_{\rm  vir} \else  $V_{\rm vir}$  \fi} 
\def \Mvir {\ifmmode M_{\rm  vir} \else $M_{\rm  vir}$ \fi}  
\def \Nvir {\ifmmode N_{\rm  vir} \else $N_{\rm  vir}$ \fi}  
\def \Jvir {\ifmmode J_{\rm vir} \else $J_{\rm vir}$ \fi} 
\def \Evir {\ifmmode E_{\rm vir} \else $E_{\rm vir}$ \fi} 
\def \vvir {\ifmmode v_{\rm vir} \else $v_{\rm vir}$ \fi} 
\def \lam {\ifmmode \lambda  \else $\lambda$ \fi} 
\def \lamp {\ifmmode \lambda^{\prime} \else $\lambda^{\prime}$  \fi} 
\def \Vmax {\ifmmode V_{\rm  max} \else  $V_{\rm max}$  \fi} 
\def \Mdm {\ifmmode M_{\rm  dm} \else $M_{\rm  dm}$ \fi}

\def \Mgas {\ifmmode M_{\rm gas} \else $M_{\rm gas}$ \fi} 
\def \Mcg {\ifmmode M_{\rm cg} \else $M_{\rm cg}$\fi} 
\def \Mhg {\ifmmode M_{\rm hg} \else $M_{\rm hg}$ \fi} 
\def \Mdisc {\ifmmode M_{\rm disc} \else $M_{\rm disc}$ \fi} 
\def \Md {\ifmmode M_{\rm d} \else $M_{\rm d}$ \fi} 
\def \Mda {\ifmmode M_{\rm d,0\%} \else $M_{\rm d,0\%}$ \fi} 
\def \Mdb {\ifmmode M_{\rm d,20\%} \else $M_{\rm d,20\%}$ \fi} 
\def \Mdc {\ifmmode M_{\rm d,40\%} \else $M_{\rm d,40\%}$ \fi} 
\def \md {\ifmmode m_{\rm d} \else $m_{\rm d}$ \fi} 
\def \Mb {\ifmmode M_{\rm b} \else $M_{\rm b}$ \fi} 
\def \Mbh {\ifmmode M_{\rm b,pri} \else $M_{\rm b,pri}$ \fi} 
\def \Mbs {\ifmmode M_{\rm b,sat} \else $M_{\rm b,sat}$ \fi} 
\def \zo {\ifmmode z_{0} \else $z_{0}$ \fi} 
\def \rd {\ifmmode r_{\rm d} \else $r_{\rm d}$ \fi}
\def \rg {\ifmmode r_{\rm g} \else $r_{\rm g}$ \fi}
\def \rb {\ifmmode r_{\rm b} \else $r_{\rm b}$\fi}
\def \rs {\ifmmode r_{\rm s} \else $r_{\rm s}$\fi}
\def \rc {\ifmmode r_{\rm c} \else $r_{\rm c}$\fi}
\def \rvir {\ifmmode r_{\rm vir} \else $r_{\rm vir}$\fi}
\def \rbh {\ifmmode r_{\rm b,pri} \else $r_{\rm b,pri}$ \fi} 
\def \rbs {\ifmmode r_{\rm b,sat} \else $r_{\rm b,sat}$ \fi} 
\def \zp {\ifmmode z_{\rm phot} \else $z_{\rm phot}$ \fi}
\def \zs {\ifmmode z_{\rm spec} \else $z_{\rm spec}$ \fi}
\def \Lya{\ensuremath{\mathrm{Ly}\alpha\ }}

\begin{document}

\title{MUSE integral-field spectroscopy towards the Frontier Fields Cluster Abell S1063: I. Data products and redshift identifications}  

\author{W.  Karman\inst{\ref{inst1}}\thanks{karman@astro.rug.nl} \and K. I. Caputi\inst{\ref{inst1}} \and C. Grillo \inst{\ref{inst2}} \and  I.~Balestra\inst{\ref{inst3}} \and P.~Rosati\inst{\ref{inst4}} \and  E. Vanzella\inst{\ref{inst5}}   \and D.~Coe\inst{\ref{inst6}} \and L. Christensen\inst{\ref{inst2}} \and A.~M.~Koekemoer\inst{\ref{inst6}} \and T. Kr\"uhler\inst{\ref{inst7}} \and M.~Lombardi\inst{\ref{inst8}} \and A.~Mercurio\inst{\ref{inst9}} \and M. Nonino\inst{\ref{inst3}} \and A. van der Wel.\inst{\ref{inst10}}}
 
\institute{ Kapteyn Astronomical Institute, University of Groningen, Postbus 800, 9700 AV Groningen, the Netherlands\label{inst1} 
\and Dark Cosmology Centre, Niels Bohr Institute,
  University of Copenhagen, Juliane Maries Vej 30, DK-2100 Copenhagen,
  Denmark\label{inst2}
\and INAF - Osservatorio Astronomico di Trieste, via
  G. B. Tiepolo 11, I-34143, Trieste, Italy \label{inst3}
\and Dipartimento di Fisica e Scienze della Terra,
  Universit\`a degli Studi di Ferrara, Via Saragat 1, I-44122 Ferrara,
  Italy \label{inst4}
\and INAF–Bologna Astronomical Observatory, via Ranzani 1, I-40127 Bologna, Italy \label{inst5}
\and Space Telescope Science Institute, 3700 San Martin
  Drive, Baltimore, MD 21208, USA \label{inst6}
\and European Southern Observatory, Alonso de Córdova 3107, Vitacura, Casilla 19001 Santiago 19, Chile \label{inst7}
\and Dipartimento di Fisica, Universit\`a degli Studi di Milano, via Celoria 16, 1-20133 Milano, Italy \label{inst8} 
\and INAF - Osservatorio Astronomico di Capodimonte, Via
  Moiariello 16, I-80131 Napoli, Italy \label{inst9}
\and Max-Planck-Institut f\"ur
 Astronomie, K\"onigstuhl 17, 69117 Heidelberg, Germany \label{inst10}
}

\abstract{
We present the first observations of the Frontier Fields Cluster Abell S1063
taken with the newly commissioned Multi Unit Spectroscopic Explorer (MUSE) 
integral field spectrograph. Because of the relatively 
large field of view (1 arcmin$^2$), MUSE is ideal to simultaneously target multiple 
galaxies in blank and cluster fields over the full optical spectrum. We analysed the four hours 
of data obtained in the Science Verification phase on this cluster
and measured redshifts for 53 galaxies. We confirm the redshift of five cluster galaxies,
and determine the redshift of 29 other cluster members. Behind the cluster, we find 17 galaxies at higher redshift,
including three previously unknown Lyman-$\alpha$ emitters at $z>3$, and five multiply-lensed galaxies. We report 
the detection of a new $z=4.113$ multiply lensed galaxy, with images that are consistent with lensing model predictions
derived for the Frontier Fields.
We detect \ion{C}{III}], \ion{C}{IV}, and \ion{He}{II} emission in a multiply lensed galaxy at $z=3.116$, suggesting
the likely presence of an active galactic nucleus. We also created narrow-band images from
 the MUSE datacube to automatically search
for additional line emitters corresponding to high-redshift candidates, but we could not identify any significant 
detections other than those found by visual inspection. With the new redshifts, it will become possible
to obtain an accurate mass reconstruction in the core of Abell S1063 through refined strong lensing
modelling.
Overall, our results illustrate the breadth of scientific topics that can be addressed with a single MUSE pointing.
We conclude that MUSE is a very efficient instrument to observe 
galaxy clusters, enabling their mass modelling, and to perform a blind search for high-redshift galaxies. 
}
              
\date{Received ... /
Accepted ...}


\titlerunning{High-z galaxies lensed by AS1063}
\authorrunning{W. Karman et al.}

\setcounter{footnote}{1}

\maketitle

\section{Introduction}
\label{sec:intro}

Spectroscopic studies of high-redshift ($z>2$) galaxies provide very valuable information on the 
mechanisms for galaxy growth and evolution in the young Universe. Multiple spectroscopic campaigns targeting high-$z$ galaxies have now been carried out, shedding light on general star formation properties \citep[e.g.][]{Shapley2003,Kajisawa2010,LeFevre2013}, chemical composition \citep[e.g.][]{Erb2006a,Maier2014}, and the presence of gas outflows \citep[e.g.][]{Erb2006b,Steidel2010,KHainline2011,Talia2012,Karman2014}. The search for Lyman-$\alpha$ emitters has also allowed the existence of multiple galaxy candidates at very high-$z$ to be confirmed \citep[$z>5$;][]{Pentericci2011,Stark2011,Curtis-Lake2012,Shibuya2012,Finkelstein2013,Guaita2013}.

A next, more challenging step is obtaining spectral information coupled to spatial information within each galaxy, through integral field spectroscopy (IFS). Integral field spectroscopy studies have been shown to be very important for revealing the dynamics of star formation in galaxies at $z\sim1-2$ \citep[e.g.][]{Forster2006,Epinat2009,Contini2012}, but only a few attempts have so far been possible at higher redshifts \citep[e.g.][]{Mannucci2009,Blanc2011}. The main reason for this limitation is that the surface brightness  of galaxies decreases with $(1+z)^4$, and thus IFS observations of high-$z$
galaxies with current telescopes require very long exposure times.  

\begin{figure*}
\begin{center}
\includegraphics[width=\textwidth]{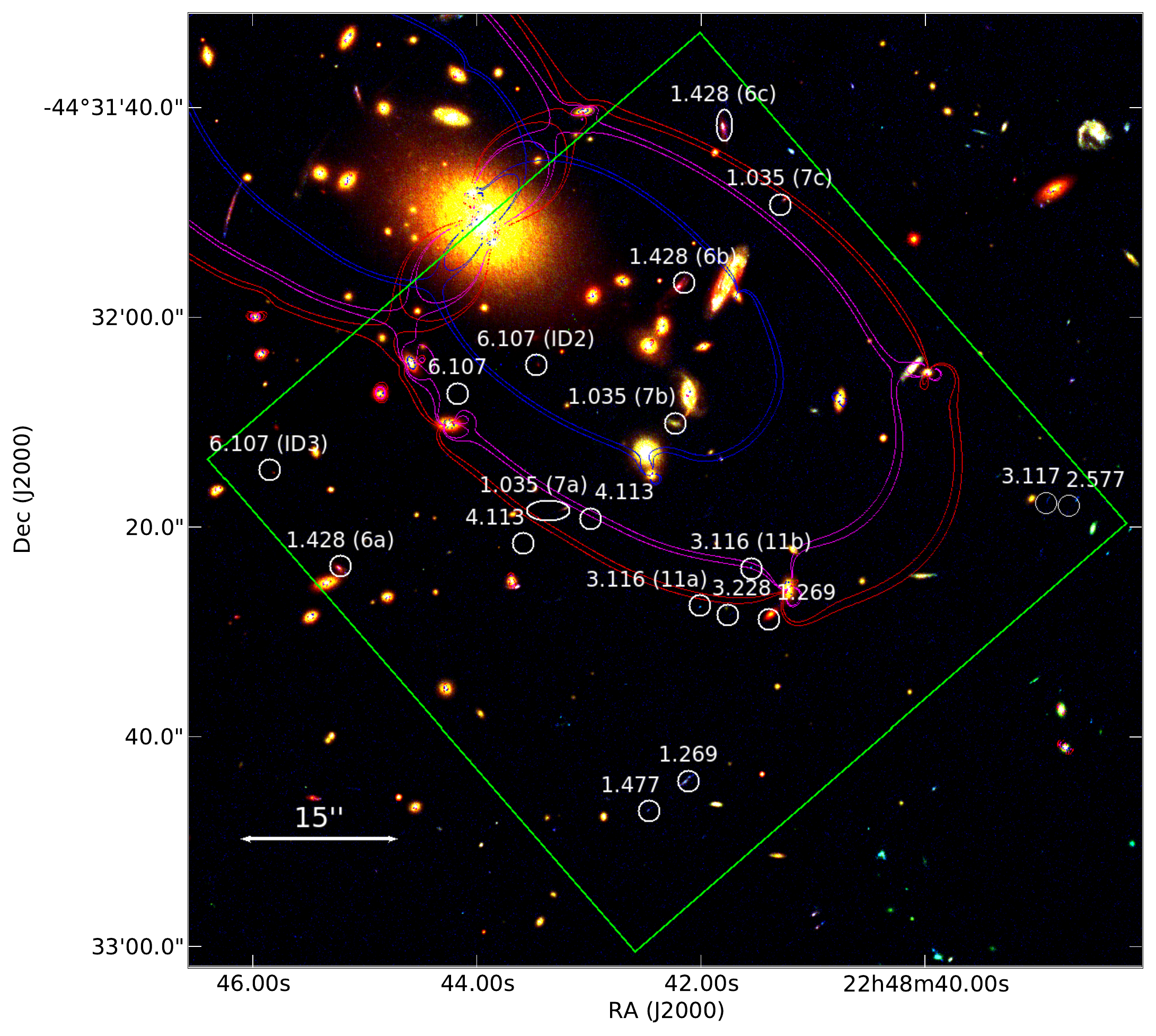}
\caption{Red-green-blue image composition of {\em HST} images of the part of AS1063 that
we observed with MUSE, where the colours in this image are
{\em blue = F435W + F475W, green = F606W + F625W + F775W + F814W + F850LP, red = F105W + F110W + F140W + F160W.}
 The individual {\em HST} images were obtained and distributed by the CLASH collaboration.
The green square delimits the 1$\times$1\arcmin\ MUSE FOV. All 19 objects with a spectroscopic redshift $z>1$ 
 are shown as white circles and their redshifts. Numbered galaxies were noted before by \citet{Monna2014} and
\citet{Balestra2013}. The blue, magenta, and red contours correspond to regions where the magnification is $\mu>200$ 
at $z=1$, 2, and 4, respectively. The magnifications are based on the second version of the Sharon models \citep{Johnson2014}, which were chosen for illustrative purposes, but we note that this is only one out of many different available models. \label{fig:fov_rgb}} 
\end{center}
\end{figure*}

Lensing cluster fields are ideal to overcome this limitation, and provide an excellent tool to investigate the high-redshift Universe.  In particular, the {\em Hubble Space Telescope (HST)} Multi-Cycle Treasury Programme CLASH 
\citep{Postman2012} has provided deep images over 16 wavebands for 25 cluster fields, enabling a series of photometric studies to identify high-$z$ galaxy candidates \citep{Jouvel2014,Monna2014,Vanzella2014}, and model in detail the lensing properties of such systems \citep{Coe2012,Zitrin2013,Grillo2014a}. A major spectroscopic follow-up of a subset of these clusters is currently in progress with the VIsible MultiObject Spectrograph (VIMOS) at the Very Large Telescope (VLT), in the so-called CLASH-VLT Large Programme (P.I.: P. Rosati), which has already yielded spectroscopic confirmation of a $z>6$ galaxy candidate \citep{Balestra2013,Richard2014a,Johnson2014}, and allowed for more accurate lensing models for some of these clusters \citep{Biviano2013,Grillo2014b}.

With the aim of investigating one of these cluster fields in further detail, 
we carried out the follow-up of Abell S1063 (RXCJ2248.7-4431) with the new Multi Unit 
Spectroscopic Explorer (MUSE) IF spectrograph on the VLT \citep{Bacon2010}.
  Abell S1063 is one of the four CLASH lensing clusters that was selected for
 ultra-deep observations as part of the {\em HST} Frontier Fields Programme
 (Lotz et al. in prep.). As a pilot programme, we obtained medium-depth
 data from MUSE over a single  pointing covering $\sim 1 \times 1$~arcmin$^2$,
 with 4~hour observations carried out in the MUSE Science Verification phase.
A main aim
of this paper is thus to assess the performance of MUSE on a number of science cases.

The layout of this paper is as follows. In Section \ref{sec:data} we give a brief
overview of the MUSE performance and the obtained data, followed by the
data reduction process. In Section \ref{sec:results}, we describe the obtained spectroscopic results, including
the determined redshifts and emission line properties. We give details of narrow-band images
 constructed from the MUSE data in Section \ref{sec:narrowband}
to search for weaker sources, and we summarise
and discuss our findings in Section \ref{sec:discussion}.


\section{Observations}
\label{sec:data}

MUSE is the newly installed integral field (IF) spectrograph on the VLT at the Paranal Observatory \citep{Bacon2012}. 
The spectrograph has a wide wavelength coverage from 4800 through 9300 \AA\ in the nominal mode, 
with an average resolution of $R\sim3000$. MUSE's field of view (FOV) is of $\sim$1~arcmin$^2$,
and therefore, MUSE combines a high spatial resolution with a large FOV, resulting in more spectral pixels (spaxels)
than any other optical IF spectrograph. 
Because of the relatively large FOV size, MUSE is very
well suited to simultaneously target multiple background galaxies in 
blank and cluster fields.
This capability is further improved by its high spatial 
resolution (0.2\arcsec\ pixel scale), which also allows for detailed 
spatial profiles of the galaxies, and is only limited by seeing until 
adaptive optics becomes available.

\begin{figure*}
\begin{center}
\includegraphics[width=\textwidth]{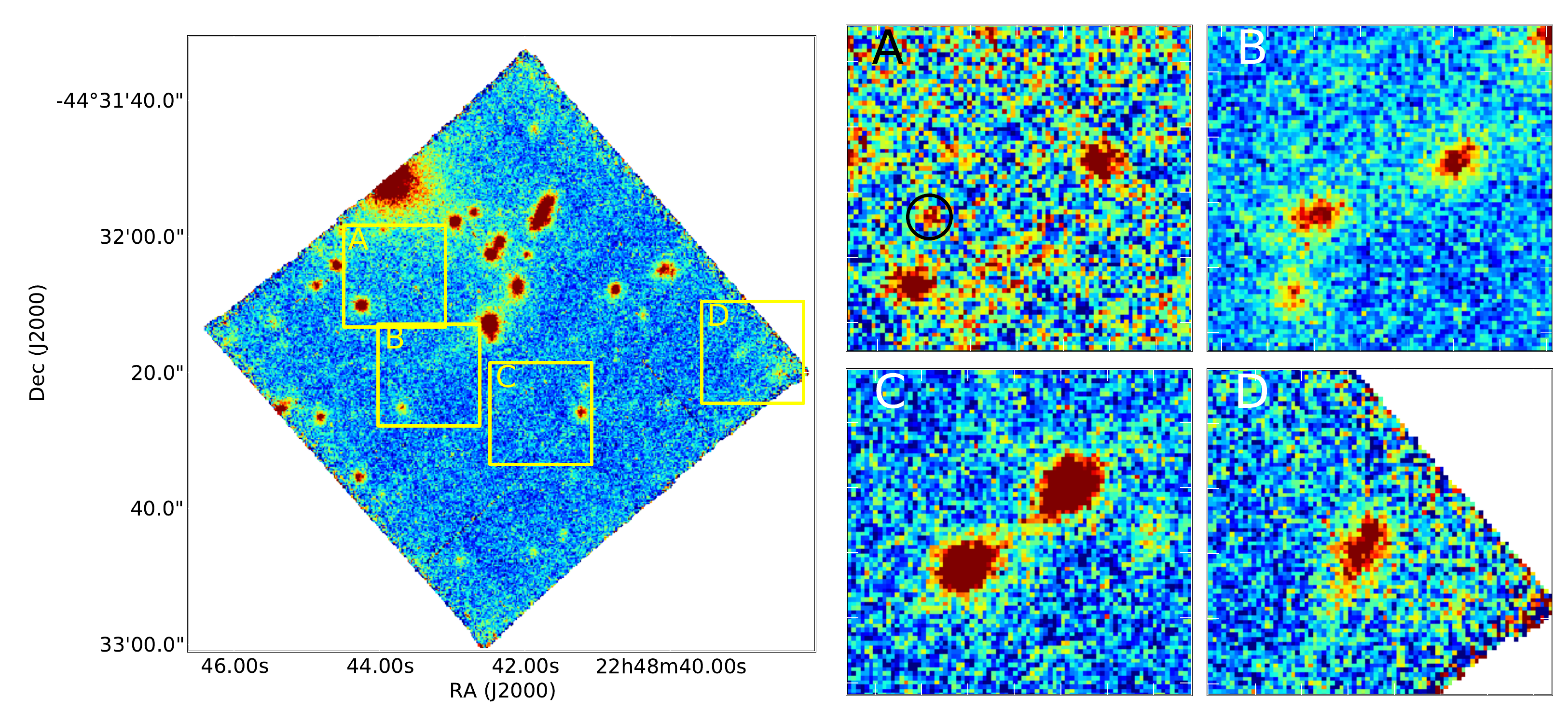}
\caption{Snapshot of our MUSE pointing towards Abell S1063.  {\em Left:} A slice of 
reduced MUSE data cube centred at 7250 \AA\ with $\Delta\lambda=1.25$ \AA.  
{\em Right:} Zoom-ins of the four regions with LAEs marked on the left panel, where the 
wavelength is centred at the peak of \Lya for galaxies 53a,c 52a,b, 49a,b, and 50 in panels A, B, C,
and D, respectively. In zoom-in A, we placed a circle around the weak 
additional detection of the quintiply lensed galaxy at $z=6.107$.\label{fig:fov_muse}} 
\end{center}
\end{figure*}

We observed the cluster Abell S1063 with a pointing centred at $\alpha=$22:48:42.217,
$\delta=$-44:32:16.850, see Fig. \ref{fig:fov_rgb}. We set the position angle to 41$^\circ$
in order to cover the area with maximum magnification according to the CLASH lensing models distributed 
for the Frontier Fields. Furthermore, this position angle and centre maximises the number of high-redshift
candidates and lensed objects as obtained from \citet{Monna2014} and the CLASH Image Pipeline.

Our MUSE observations were carried out as part of the Science Verification phase of MUSE in the nights of
June 25th 2014 ($2\times1420$ s) and June 29th 2014 ($6\times1420$ s). The 1 hour of observations on the 25th were 
carried out with a seeing of 1\arcsec, while the 3 hours of observations on the 29th had a seeing of 1.1\arcsec. 
We applied a dither pattern consisting of a positional offset and a rotation in our observation strategy to 
better remove cosmic rays and to obtain a better noise map. The positional offset consisted of a shift of the
centre by 0.4-0.8\arcsec, so that every exposure had a slightly different centre. In addition, we rotated
every exposure by 90$^\circ$, resulting in two exposures for every position angle.

\subsection{Data reduction}

We reduced the raw data using the MUSE Data Reduction Software version 0.18.2, which contains
all standard reduction procedures. We show a slice of the resulting datacube in Fig. \ref{fig:fov_muse}. 
The pipeline subtracts the bias, applies a flatfielding and calibrates the wavelength
by using more than 70 arc lamp lines distributed over the whole wavelength range. The wavelength solution is recalibrated 
using the atmospheric sky lines in the science exposures. We checked the wavelength and tracing solutions
for every IFU and slice, and verified that the residuals were typically $<0.15$ \AA. For two out of the eight exposures, we removed
one slice from one IFU (15\arcsec$\times$0.2\arcsec\ spatially) as the detector did not work well in this slice for
 temperatures below 7 degrees Celsius, and the tracing and flatfielding recipes fail. This is consistent with findings of
preliminary MUSE testing (E. Valenti, private communication), and has now been corrected.
As a last step, the pipeline calibrates the flux in the exposures using a standard star, and combines all exposures
into one 3D datacube, where differential atmospheric diffraction is taken into account. Although the pipeline corrects the data for a small number of 
telluric lines, many atmospheric lines are still present in the data. We masked the strongest lines
in the data manually, but kept the weaker ones, as the data at these wavelengths can still provide 
useful information when observing emission lines.

To remove cosmic rays from the data, we applied the {\sc python} routine of {\sc LA Cosmic} \citep{Dokkum2001} to the
raw science data frames, in addition to the cosmic ray removal available in the pipeline.
Because the sky is not completely removed from the datacube, we created a mask consisting of eleven regions free
of sources, and subtracted the mean of these regions at each wavelength. We identified sources from this 
sky-subtracted datacube by careful visual inspection, and used the {\sc Common Astronomical Software Application (CASA)} package\footnote{http://casa.nrao.edu} 
to collapse the data in the spatial directions within a fitted ellipse and extract 1D spectra. 
The astrometry was assigned using the distributed astrometry table.
To test the positional accuracy, we used the {\em HST} images for Abell S1063 as reference.
We created three narrowband images with the
MUSE observations using the pipeline, and ran {\sc sextractor} \citep{Bertin1996} on both sets of images. We found
a median positional accuracy of 0.12$\arcsec$ for MUSE compared to {\em HST}.

\citet{Fumagalli2014} found that they needed to rescale the illumination in every slice
to properly reduce the data. In addition, they manually created a sky model to subtract
from the intermediately produced so-called {\sc pixtables} to improve the signal.
We developed {\sc python} routines to perform similar steps in our data reduction,
but we do not find a substantial improvement to the quality of our data. Therefore,
we do not use the data reduced with these illumination or sky corrections in this paper. 

Figure \ref{fig:fov_muse} shows the reduced MUSE data at 7250 \AA. A large number of sources 
and the overall good data quality are clearly evident from Fig. \ref{fig:fov_muse}.

\section{Spectral analysis}
\label{sec:results}

We extracted 1D spectra for 61 detections, belonging to 53 individual galaxies, in our observed 1 arcmin$^2$ field, and determined 
their redshifts, see Tables \ref{tab:results1} and \ref{tab:results2}. Of these
objects, 34 are galaxies within 5000 \kms of the cluster redshift ($z_{\rm c}=0.348$), 25 
detections correspond to 17 individual galaxies at higher
redshift, and 2 are foreground galaxies. We find five more objects with spectral features that are too weak to 
securely determine a redshift.

\subsection{Cluster galaxies}

\begin{figure}
\begin{center}
\includegraphics[width=\columnwidth]{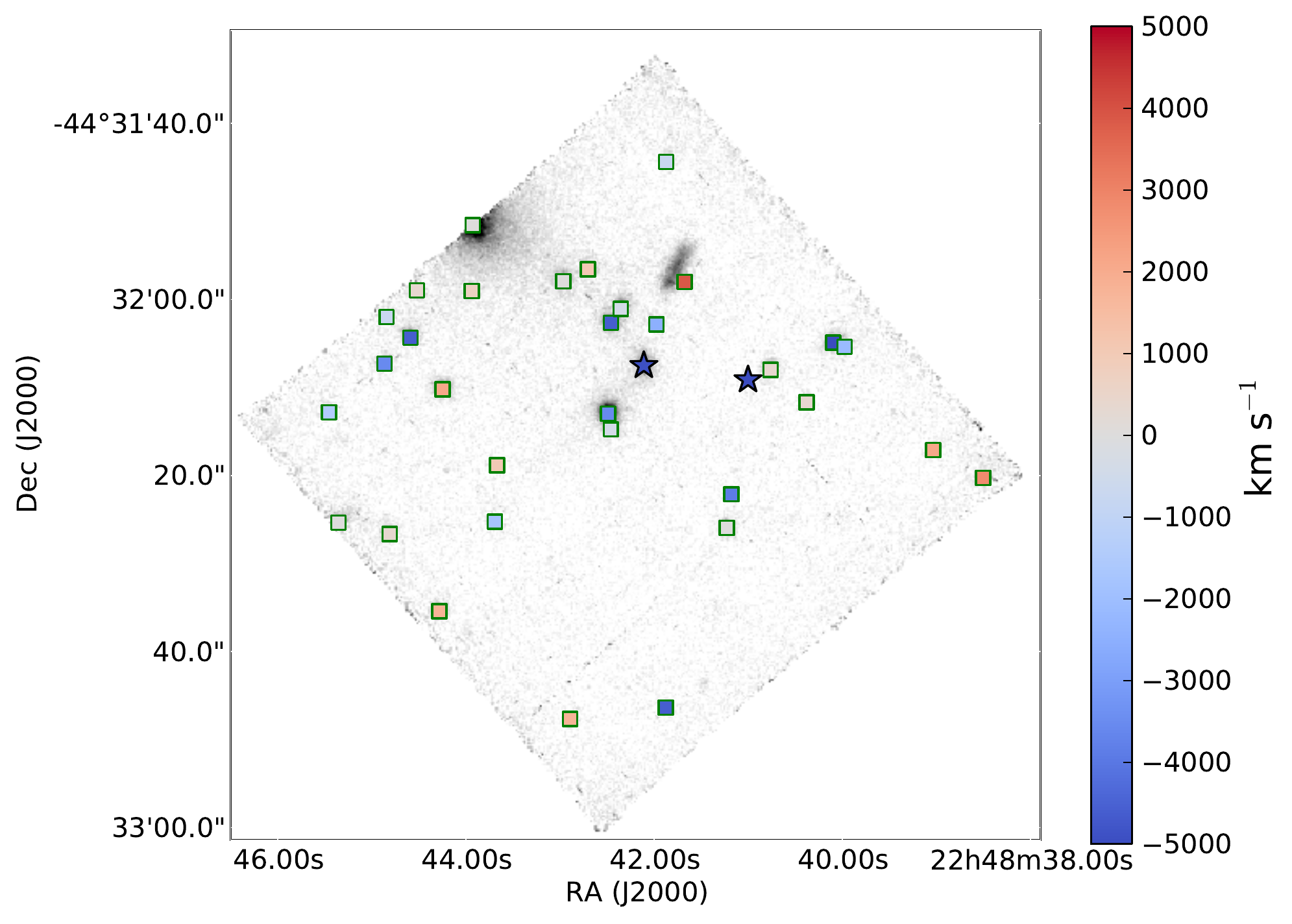}
\caption{Distribution map of the identified cluster galaxies. Squares correspond to passive galaxies,
while stars indicate active galaxies, where their classification is based on the presence
or absence of optical emission lines. The galaxies have been coloured according to their velocity
relative to the cluster, with bluer colours meaning higher velocities towards us, and redder colours 
corresponding to higher velocities away from us.\label{fig:fov_act}}
\end{center}
\end{figure}

\begin{table*}
\begin{center}
 \begin{tabular}{cccc|cc|ccc|c}
 {\bf ID}& {\bf RA} (J2000) & {\bf DEC} (J2000) & $F775W$ & $z$ & {\bf QF} & $z_{\rm prev.}$ & ID$_{\rm ref.}$ & $z_{{\rm VIMOS}}$& {\bf Active/Passive}\\
  \hline 
1  & 22:48:40.098 & -44:32:04.90 & 20.81 & 0.327 & 1 & - & - & 0.327 & P \\
2  & 22:48:42.458 & -44:32:02.67 & 19.84 & 0.335 & 1 & 0.334 & 835 & 0.333 & P\\
3  & 22:48:44.590 & -44:32:04.34 & -     & 0.335 & 1 & - & - & - & P \\
4  & 22:48:41.877 & -44:32:46.37 & 22.32 & 0.335 & 1 & - & - & 0.335 & P \\
5  & 22:48:41.003 & -44:32:09.13 & 23.14 & 0.336 & 1 & - & - & 0.336 & A \\
6  & 22:48:42.110 & -44:32:07.51 & 19.15 & 0.337 & 1 & 0.336 & 779  & 0.336 & A\\
7  & 22:48:41.179 & -44:32:22.14 & 22.37 & 0.337 & 2 & - & - & - & P\\
8  & 22:48:44.865 & -44:32:07.31 & 20.93 & 0.338 & 1 & - & - &0.338 & P\\
9  & 22:48:42.489 & -44:32:12.99 & 19.04 & 0.338 & 1 & 0.336 & 746 & 0.337 & P(A)\\
10 & 22:48:41.975 & -44:32:02.83 & 21.36 & 0.341 & 1 & - & - & - & P\\
11 & 22:48:39.977 & -44:32:05.38 & 21.50 & 0.342 & 1 & - & - & - & P\\
12 & 22:48:43.692 & -44:32:25.24 & 21.58 & 0.343 & 1 & - & - & 0.344 & P \\
13 & 22:48:45.456 & -44:32:12.82 & 22.00 & 0.344 & 1 & - & - & 0.343 & P \\
14 & 22:48:41.872 & -44:31:44.37 & 22.11 & 0.346 & 1 & - & - & - & P\\
15 & 22:48:44.843 & -44:32:02.01 & -     & 0.346 & 2 & - & - & - & P\\
16 & 22:48:42.355 & -44:32:01.06 & 20.23 & 0.347 & 1 & 0.346 & 850 & 0.347 & P \\
17 & 22:48:42.461 & -44:32:14.77 & 21.33 & 0.347 & 2 & - & - & - & P\\
18 & 22:48:43.927 & -44:31:51.55 & 17.09 & 0.348 & 2 & - & - & 0.347 & P \\
19 & 22:48:44.521 & -44:31:58.98 & 22.02 & 0.348 & 1 & - & - & - & P\\
20 & 22:48:45.354 & -44:32:25.35 & 20.47 & 0.348 & 1 & 0.348 & 575 & - & P\\
21 & 22:48:41.228 & -44:32:25.95 & 20.78 & 0.348 & 1 & - & - & - & P\\
22 & 22:48:42.967 & -44:31:57.95 & -     & 0.349 & 2 & - & - & - & P \\
23 & 22:48:40.764 & -44:32:07.99 & 20.69 & 0.349 & 1 & - & - & 0.346 & P \\
24 & 22:48:40.382 & -44:32:11.69 & 22.21 & 0.349 & 1 & - & - & - & P\\
25 & 22:48:44.810 & -44:32:26.65 & 21.11 & 0.349 & 1 & - & - & 0.348 & P\\
26 & 22:48:43.937 & -44:31:59.05 & 20.53 & 0.350 & 1 & - & - & - &  P\\
27 & 22:48:42.703 & -44:31:56.59 & 20.61 & 0.351 & 1 & - & - & - & P\\
28 & 22:48:43.670 & -44:32:18.85 & 22.89 & 0.351 & 2 & - & - & - & P\\
29 & 22:48:44.284 & -44:32:35.44 & 20.98 & 0.353 & 1 & - & - & - & P\\
30 & 22:48:42.893 & -44:32:47.66 & 22.19 & 0.353 & 2 & - & - & 0.349 & P \\
31 & 22:48:44.247 & -44:32:10.22 & -     & 0.354 & 1 & - & - & 0.353 & P\\
32 & 22:48:39.036 & -44:32:17.12 & 22.53 & 0.355 & 2 & - & - & - & P\\
33 & 22:48:38.505 & -44:32:20.27 & 22.11 & 0.356 & 2 & - & - & - & P\\
34 & 22:48:41.678 & -44:31:58.00 & 21.67 & 0.359 & 1 & - & - & - & P
 \end{tabular}
\caption{Properties of the detected cluster galaxies in the observed field. 
The coordinates in columns two and three are measured in {\sc CASA} by fitting a 2D Gaussian 
to their spatial profile at 7250 \AA\ for cluster galaxies, while
for the other galaxies the wavelength of emission lines is used. In
column four we present the isophotal magnitude in the $F775W$-band from {\em HST}
as determined by \citet{Postman2012}, where we matched the locations
using the brightest source within 1\arcsec.
The fifth column gives the MUSE redshift measured here, with a quality flag (QF) in the sixth column.
The quality flag of the redshift shows how accurately the redshift is measured, {\em 1} 
represents a very certain redshift with multiple lines that give $\Delta z<0.001$, {\em 2} 
represents redshift where several lines are detected, but because of a lower signal-to-noise ratio
or intrinsically wider lines, the redshift varies slightly between different lines ($\Delta z<0.003$).
The seventh and eighth columns contain previous redshift determination by \citet{Gomez2012},
with their previous IDs.
The ninth column quotes redshifts measured with VIMOS from Balestra et al. (in prep.).
In the last column we show the classification as active (A)
or passive (P) of each galaxy based on the presence of 
emission lines indicative of activity.\label{tab:results1}} 
\end{center}
\end{table*}

We show the properties of the cluster galaxies in Table \ref{tab:results1}. We note
that only five of the determined cluster galaxies
have a previously published redshift determination \citep{Gomez2012}, and therefore we add 
redshifts to 29 galaxies of the cluster. The redshift measured for the cluster galaxies is in
excellent agreement with measurements from VIMOS (Balestra et al., in prep).
We find that almost all the cluster galaxies have very similar spectra. In all 34 galaxies we detect
\ion{Ca}{II} H and K, and 33 show a strong 4000~\AA\ break. We used the presence or
absence of the [\ion{O}{II}]~$\lambda\lambda~3726.0,3728.8$~\AA, \ion{H}{$\beta$}, 
[\ion{O}{III}]~$\lambda~5006.8$~\AA, and \ion{H}{$\alpha$} to classify the
cluster galaxies as active if we find at least one these lines with a negative equivalent width,
 i.e. it is in emission, or passive otherwise. Out of the 34 galaxies, only two
are classified as active, while the remaining 32 are all lacking signatures
of activity. The classification active does not distinguish between star formation activity and AGN, as 
we do not determine the ionizing source.
 We show the distribution of the cluster galaxies in Fig. \ref{fig:fov_act}
in which we also indicate their classification. It is clear from this figure that
the number density of cluster members increases towards the centre of the cluster, and that almost all
bright objects with a visible continuum are part of the cluster. The two
active galaxies do not appear to be at the cluster outer regions,
as would be expected if they were recently accreted by the cluster. However,
because this only involves two galaxies, this could well be due to projection effects.

\begin{figure*}
\begin{center}
\includegraphics[width=\textwidth]{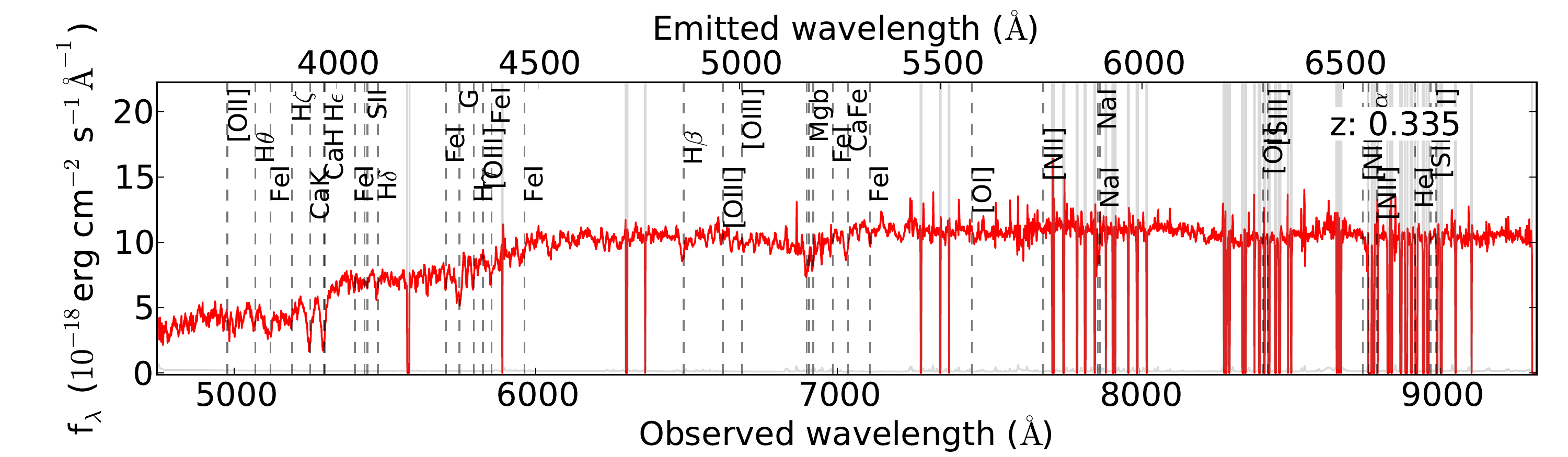}
\caption{Spectrum of a typical cluster galaxy (object 2). 
The red line represents the 1D spectrum,
where the source is extracted as an ellipse with typically 1\arcsec$\times$1\arcsec\ 
 axes and then summed in the
spatial direction. The grey line represents the level of dispersion
in the global sky, while the black dashed vertical lines represent
restframe wavelengths of the most important transitions. At skylines
grey vertical lines are plotted, and the redshift
is shown in the top right corner.\label{fig:bcg1}}
\end{center}
\end{figure*}

\begin{figure*}
\begin{center}
\includegraphics[width=\textwidth]{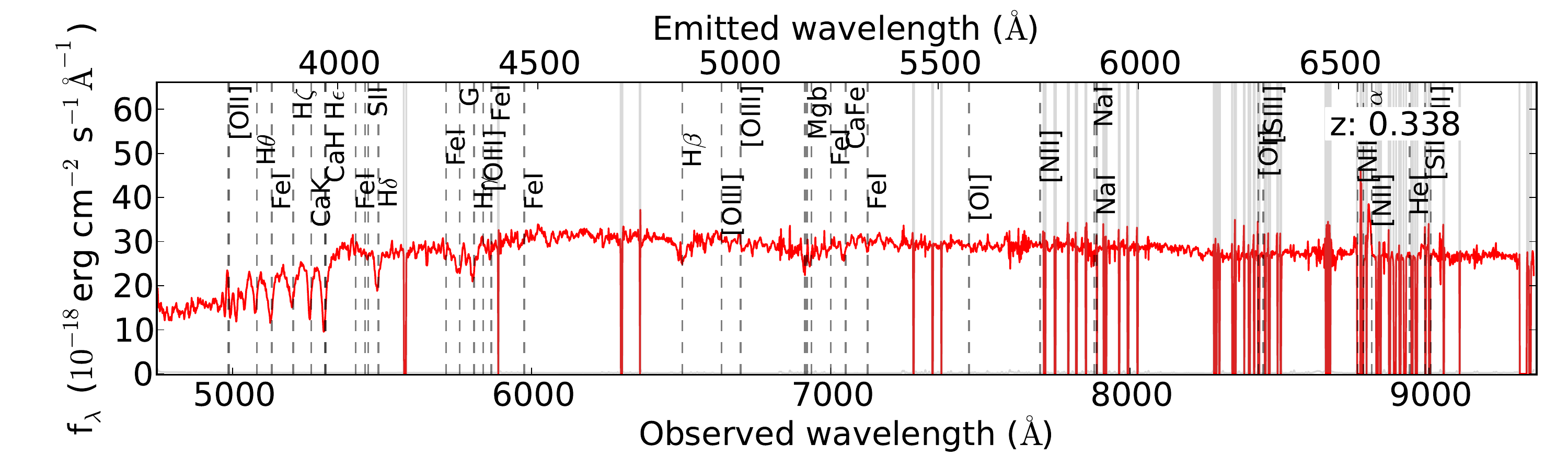}
\caption{Spectrum of a second example of a
cluster galaxy (object 9). Lines and legends are the same as in Fig.
\ref{fig:bcg1}.\label{fig:bcg2}}
\end{center}
\end{figure*}

In Figs. \ref{fig:bcg1} and \ref{fig:bcg2} we show two example spectra of bright cluster galaxies to
illustrate the spectral properties. Figure \ref{fig:bcg1} shows the spectrum of object 2.
We see a bright continuum, with two strong absorption 
lines between 5220 and 5320 \AA, identified as the \ion{Ca}{II} H and K lines.
Furthermore, we see a clear absorption feature at the {\em G}-band, and all
Balmer lines are in absorption. Absorption of the \ion{Mg}{I} line at 5175 \AA\ rest frame
also indicates an old stellar population and is clearly visible in the spectrum.
In agreement with the purely old population, there are no emission lines visible
over the whole wavelength range. 

Object 9, Fig. \ref{fig:bcg2}, shows strong absorption lines around
4000~\AA, easily recognisable as the \ion{Ca}{II} and \ion{H}{$\delta$} lines. The 4000~\AA\ 
break is clear, and we see a small but broad absorption corresponding to
\ion{H}{$\beta$}. Although all of these properties indicate
that the stellar population is old, we find some [\ion{O}{II}]~$\lambda\lambda~3726.0,3728.8$~\AA\
and clear [\ion{N}{II}]~$\lambda6585$ \AA\ 
emission. Because there is no [\ion{O}{III}] emission visible, it is unlikely
that an AGN is responsible for the [\ion{O}{II}] emission, as AGN have a higher ionisation level. 
Post-starburst galaxies sometimes show weak [\ion{O}{II}] emission, however, 
the \ion{H}{$\delta$} equivalent width measured for this object, $EW\approx 0.5$ \AA, 
is significantly less than the $EW=3-5$~\AA\ observed in post-starburst galaxies.
Therefore, we conclude that this emission of [\ion{O}{II}] indicates a young
stellar population coexisting with the underlying old stellar population. 

\begin{figure*}
\begin{center}
\includegraphics[width=\textwidth]{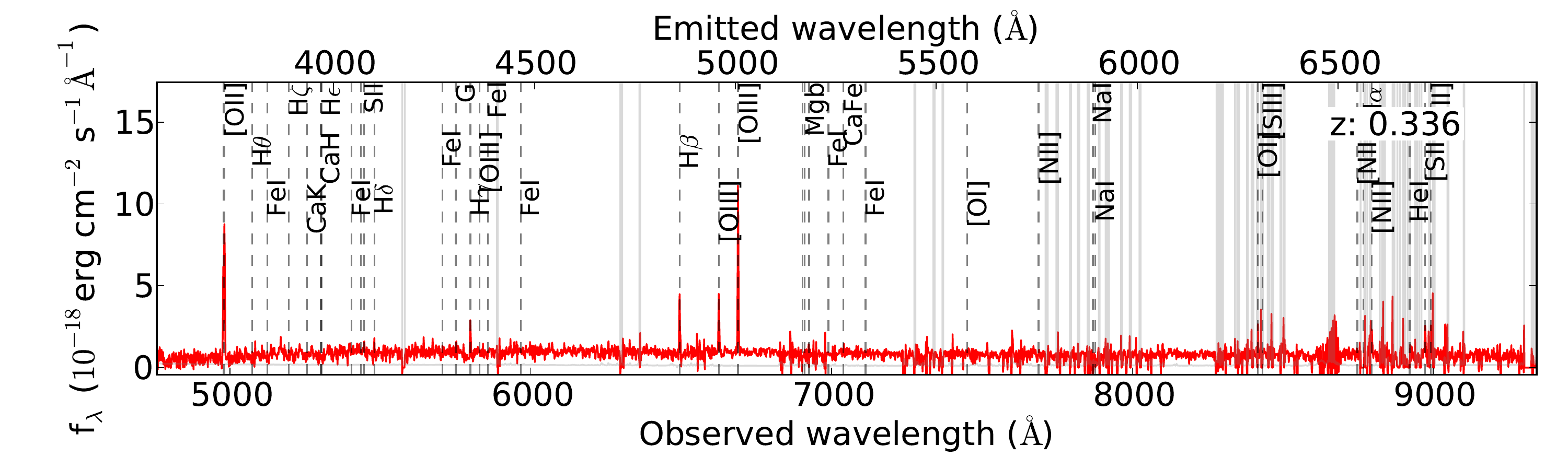}
\caption{Spectrum of object 5, one of the two cluster galaxies classified
in this work as active, based on its (clearly visible) emission lines.
Lines and legends are the same as in Fig.
\ref{fig:bcg1}.\label{fig:ceg1}}
\end{center}
\end{figure*}

We show the spectrum of one cluster galaxy (object 5) classified as active
in Fig. \ref{fig:ceg1}. There is no clear continuum recognised by visual inspection for this source,
but we do see several emission lines, including [\ion{O}{II}], \ion{H}{$\gamma$},
\ion{H}{$\beta$}, and [\ion{O}{III}]. The only strong emission
line that we do not detect is \ion{H}{$\alpha$}, as it is hidden behind
a skyline. However, we see a weak signal at the red side of the skyline,
indicating that the \ion{H}{$\alpha$} line may also be strong.  When spatially collapsing the data,
we see that a continuum is visible, including the strongest absorption features.

\subsection{Background galaxies}

We identified many other sources that are not part of AS1063. Only
two sources are found to have a redshift placing them in front of 
the cluster. These foreground galaxies are identified on a 
set of hydrogen emission lines (e.g. \ion{H}{$\alpha$} and \ion{H}{$\beta$}).

\subsubsection{Intermediate {\em z} galaxies}

\begin{table*}
\begin{center}
 \begin{tabular}{cccc|cc|cc}
 {\bf ID}& {\bf RA} (J2000) & {\bf DEC} (J2000) & $F775W$ & $z$ & {\bf QF} & $z_{prev}$ & ID$_{\rm ref.}$\\
  \hline 
35  & 22:48:43.394 & -44:32:47.28 & 23.33 & 0.153 & 1 & - & - \\
36  & 22:48:41.195 & -44:31:44.24 & 23.47 & 0.160 & 1 & - & - \\ 
37  & 22:48:39.980 & -44:32:24.75 & 22.14 & 0.459 & 1 & - & - \\
38  & 22:48:41.767 & -44:31:56.80 & 18.79 & 0.607 & 1 & 0.610 & 878 (3)\\
39  & 22:48:45.288 & -44:32:16.38 & 22.86 & 0.611 & 1 & - & - \\
40  & 22:48:44.260 & -44:32:22.46 & 24.09 & 0.652 & 1 & - & - \\
41  & 22:48:43.019 & -44:32:31.90 & 26.73 & 0.698 & 1 & - & - \\
42  & 22:48:43.233 & -44:32:38.65 & 25.31 & 0.714 & 1 & - &  667 (3) \\
43a & 22:48:41.302 & -44:31:49.86 & 26.21 & 1.035 & 1 & - & 7a (2)\\
43b & 22:48:42.234 & -44:32:10.13 & -     & 1.035 & 1 & - & 7b (2)\\ 
43c & 22:48:43.366 & -44:32:19.16 & -     & 1.035 & 1 & - & 7c (2)\\
44  & 22:48:41.407 & -44:32:28.61 & 23.09 & 1.269 & 2 & 1.270  & (1) \\
45  & 22:48:42.125 & -44:32:44.08 & 23.77 & 1.269 & 9 & 1.269 & A (6) \\
46a & 22:48:45.223 & -44:32:23.88 & 22.05 & 1.428 & 1 & 1.429 & 6a/2.1 (1,2,5,6)\\
46b & 22:48:42.199 & -44:31:57.29 & -    &  1.428 & 1 & 1.429 & 6b/2.2 (2,5)\\
46c & 22:48:41.781 & -44:31:42.39 & 22.63 & 1.428 & 1 & 1.429 & 6c/2.3 (2,5,6) \\
47  & 22:48:42.477 & -44:32:46.77 & 24.85 & 1.477 & 9 & - & - \\
48  & 22:48:38.723 & -44:32:18.05 & - & 2.577 & 9 & 2.576 & (4) \\
49a & 22:48:42.011 & -44:32:27.69 & 25.50 & 3.116 & 1 & 3.117 & 11.1 (2,6)\\
49b & 22:48:41.564 & -44:32:24.01 & 25.33 & 3.116 & 1 & 3.110 & 11b (1,2)\\
50  & 22:48:38.939 & -44:32:17.85 & 24.14& 3.117 & 1 & - & - \\
51  & 22:48:41.764 & -44:32:28.47 & 25.72& 3.228 & 1 & - & - \\
52a & 22:48:42.984 & -44:32:19.08 & 26.92 & 4.113 & 1 & - & - \\
52b & 22:48:43.593 & -44:32:21.71 & 26.32 & 4.113 & 1 & - & - \\
53a & 22:48:43.449 & -44:32:04.56 & - & 6.107 & 1 & 6.107-6.110 & ID2/6.2 (1,2,5,6) \\
53b & 22:48:45.827 & -44:32:14.62 & - & 6.107 & 1 & 6.107-6.110 & ID3/6.3 (1,2,5,6)\\
53c & 22:48:44.177 & -44:32:07.14 & - & 6.107 & 9 & - & - \\
\hline
54 & 22:48:44.579 & -44.32.18.78 & 23.35 & ? & - & - \\
55 & 22:48:40.581 & -44:32:25.44 & 23.03 & ? & - & - \\
56 & 22:48:41.320 & -44:32:35.30 & 23.24 & ? & - & - \\
57 & 22:48:43.978 & -44:32:37.88 & 22.54 & ? & - & - \\
58 & 22:48:41.465 & -44:32:43.63 & 22.91 & ? & - & - \\
\hline
60 & 22:48:43.184 & -44:32:08.16 & - & ? & - & - \\
61 & 22:48:41.908 & -44:32:13.18 & - & ? & - & - \\
62 & 22:48:43.673 & -44:32:18.89 & 22.89 & ? & - & - \\
63 & 22:48:44.371 & -44:32:26.35 & 23.27 & ? & - & - \\
64 & 22:48:44.534 & -44:32:18.92 & 23.35 & ? & - & - 
 \end{tabular}
\caption{Properties of the detected background or foreground galaxies in the observed field. 
The columns are the same as the first eight columns in Table \ref{tab:results1}, but the redshifts
and IDs in the seventh and eighth column refer to different studies (1=\citet{Balestra2013}, 2=\citet{Monna2014}, 3=\citet{Gomez2012}, 4=Balestra et al., in prep, 5=\citet{Richard2014a}, 6=\citet{Johnson2014})).
We added a quality flag in column six:
 {\em 9}~means that the redshift is based on a single emission line, although for [\ion{O}{II}] the
profile is also taken into account.
Galaxies with IDs 54-58 show flat and featureless continuums, and are identified
in the 3D datacube. Because of the featureless continuum, we are unable to determine
the redshift for these galaxies.
Sources with IDs higher than 60 are only seen in several
narrowband images created with the MUSE data and are not found in the 3D datacube. Therefore,
they do not have an extracted spectrum or redshift determination.\label{tab:results2}} 
\end{center}
\end{table*}
We identified 17 galaxies behind the cluster. A total of 11 out
of these 17 galaxies are intermediate redshift sources ($0.4<z<1.5$),
and we clearly detect [\ion{O}{II}]~$\lambda\lambda 3726.0,3728.8$~\AA\ in emission
for each of these 11 galaxies. When [\ion{O}{III}] $\lambda4959.0$~\AA\ or [\ion{O}{III}]~$\lambda5006.8$~\AA\ 
fall inside the MUSE wavelength range (six galaxies), we also identified these emission lines.
Only five out of these intermediate redshift galaxies show a continuum, while
for the other six only emission lines are seen.

\begin{figure}
 \includegraphics[width=0.9\columnwidth]{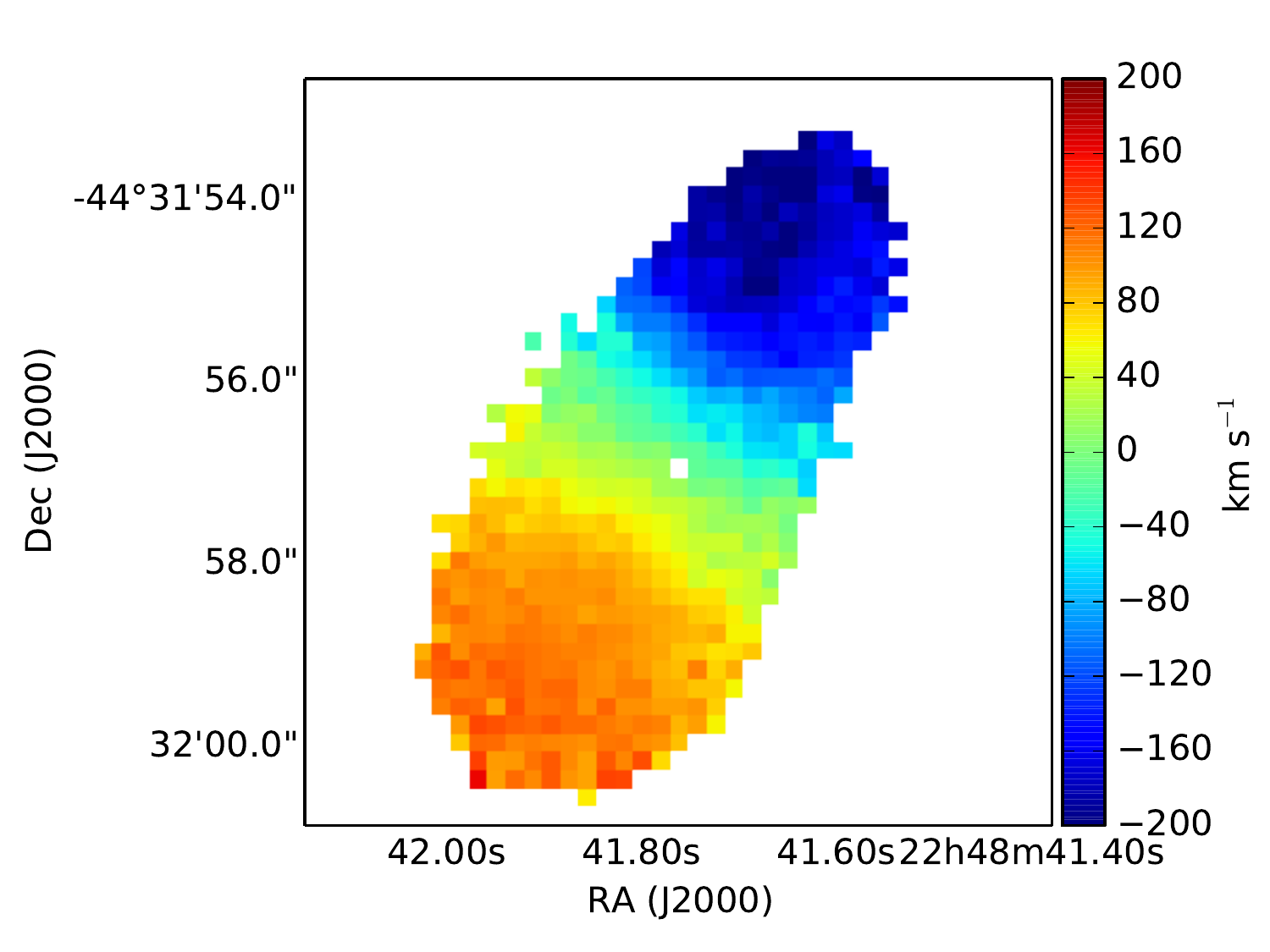}
 \caption{[\ion{O}{II}] velocity map of galaxy 38 at $z=0.607$. The 
 coordinates are shown on the axes, and the velocity 
 is given for every pixel by its colour. This figure
clearly illustrates the power of MUSE for 2D spectroscopy.\label{fig:rot1}}
\end{figure}

At intermediate and high redshifts, MUSE is well suited to study 
the kinematic profile of galaxies. To illustrate the power of
MUSE at this, we created a velocity map for
one of the intermediate redshift galaxies, object 38 in Table \ref{tab:results2} at $z=0.607$.
This galaxy has a disk-like morphology and a clear continuum with strong
emission lines. The large size of this galaxy, $\sim$5\arcsec\ or $\sim$33 kpc, 
and its edge-on orientation make it ideal to illustrate how MUSE can resolve the rotation pattern.
We fit a Gaussian to the [\ion{O}{II}] emission line profile at every
spatial location, and determine the velocity offset with respect
to the central spaxels. We show the resulting velocity map
in Fig. \ref{fig:rot1}. The galaxy shows symmetrical
rotation and the velocity reaches $\sim 200$~\kms at the largest radii.
Velocity maps of other emission lines like [\ion{O}{III}] and
\ion{H}{$\beta$} show similar maps, but their analysis is hindered
by the presence of skylines.

Sixteen lensed images, belonging to seven individual galaxies, that 
were found by \citet{Monna2014} are situated within our field.
We identified ten out of these sixteen images, and determined their
redshifts. Two triple-lensed galaxies are at intermediate redshifts,
which we identified as sources 6 and 7 from
\citet{Monna2014} and our sources 46 and 43, respectively, and 
four images correspond to galaxies at higher redshift, see Sect. \ref{sec:highz}. Galaxy 43
is at a redshift of $z=1.035$, independently obtained from all detections
 43a, 43b, and 43c.  In addition, we detect source 46 at three different places,
and we measured a redshift of $z=1.428$ at each location. This is 
consistent with the redshift reported by \citet[][source 6a, $z=1.429$]{Balestra2013},
\citet[][source 2, $z=1.429$]{Richard2014a}, and \citet{Johnson2014}.

\subsubsection{High-redshift galaxies}
\label{sec:highz}

We identified one galaxy (source 48) in the redshift gap $1.5<z<2.9$ where neither [\ion{O}{II}]
or \Lya are redshifted into the MUSE wavelength range. The redshift measurement of $z=2.577$ is based on 
\ion{C}{III}]~$\lambda~1907$~\AA, and secure thanks to a previous redshift determination 
with VIMOS (Balestra et al., in prep). A skyline prevents the detection of \ion{C}{III}]~$\lambda~1909$~\AA,
and no other emission lines are found for this source.

\begin{figure}
\begin{center}
\includegraphics[width=\columnwidth]{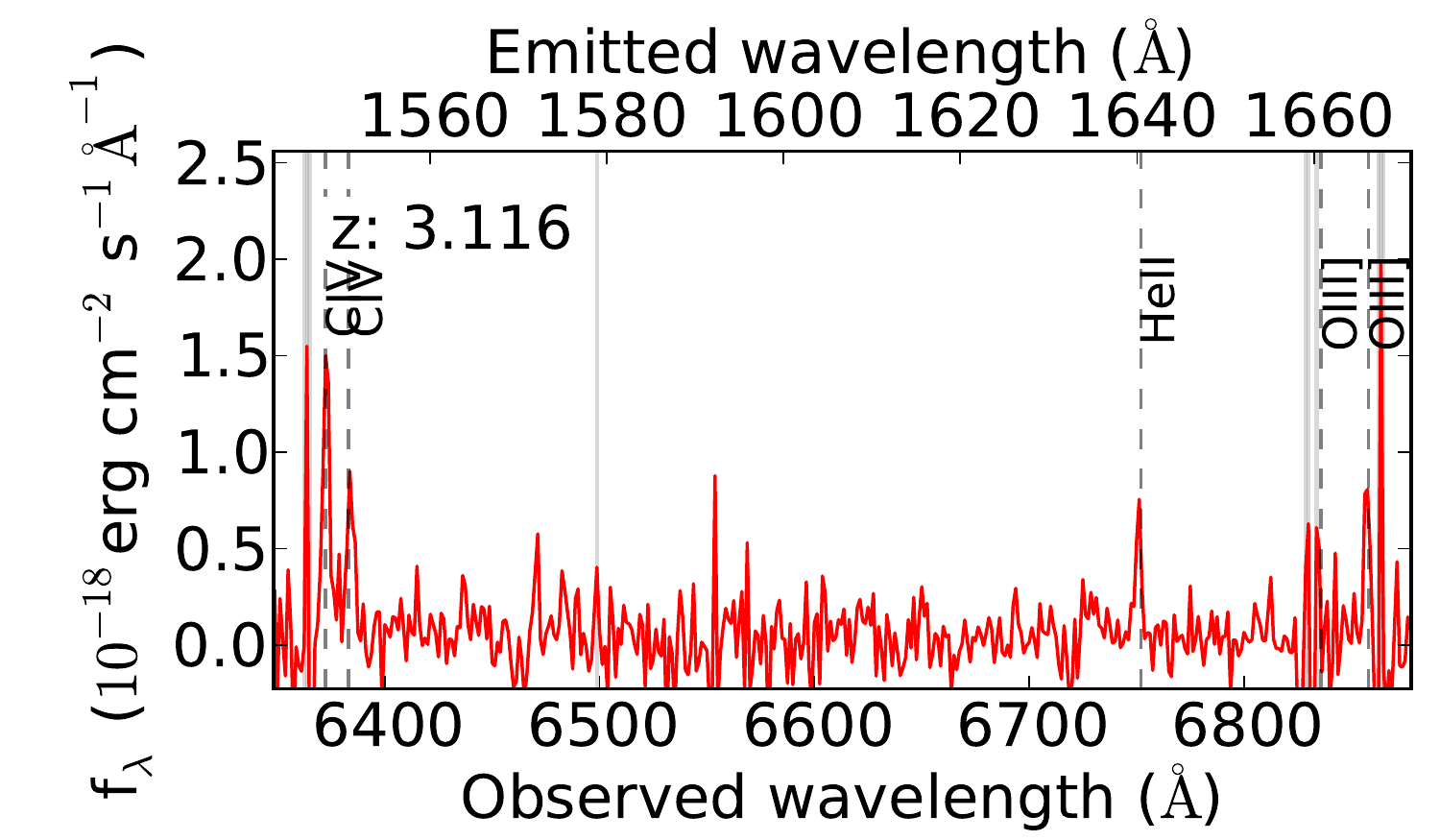}
\caption{MUSE 1D spectrum of source 49b, with wavelength on the horizontal axis and the flux on the
vertical axis. The spectrum is zoomed in on the emission lines identified as \ion{C}{IV}, \ion{He}{II},
and \ion{O}{III}]. Lines and legends are the same as in Fig. \ref{fig:bcg1}.
\label{fig:11b}} 
\end{center}
\end{figure}

At redshift $z\sim3$, we found two other sources (49a\&b) previously recognised by 
\citet{Monna2014} as a lensed galaxy (their sources 11a and 11b, 
see Fig. \ref{fig:fov_muse}c for a zoom in with MUSE),
and we measured a redshift of $z=3.116$ for them. 
This redshift is consistent with the result found by \citet{Balestra2013} and \citet{Johnson2014},
and we find additional emission lines at 6372.5, 6383.4, 6750, and 6857~\AA. 
We identified these lines as \ion{C}{IV}~$\lambda\lambda 1548,1550$~\AA, 
\ion{He}{II}~$\lambda 1640$, \ion{O}{III}]~$\lambda 1666$, and 
\ion{C}{III}]~$\lambda 1907$~\AA, see Fig. \ref{fig:11b}. Clear detections of
\ion{O}{III}]~$\lambda 1661$ and \ion{C}{III}]~$\lambda 1909$ \AA\ are not seen in the datacube,
because they are largely covered by a skyline. The redshift that we measured from all of these 
emission lines is within $\Delta z<0.001$ of $z=3.116$. The emission lines suggest AGN activity in this
galaxy \citep[e.g.][]{KHainline2011}, however, we do not detect any \ion{N}{V} emission commonly seen in AGN. 
We compared the flux ratios to those found in the Lynx 
arc \citep{Fosbury2003,Binette2003} and the M2031 arc at $z=3.51$ \citep{Christensen2012b}. 
We assumed that the ratio of the \ion{C}{III}] and \ion{C}{IV}
doublets are equal, in order to estimate the flux of \ion{C}{III}] $\lambda 1909$. The
ratio of \ion{C}{IV}/\ion{C}{III}]=3.4, which is much higher than in the Lynx arc, suggesting
AGN activity. The flux ratio of \ion{C}{IV}/\ion{He}{II} = 5 is significantly lower than for a hot star cluster,
supporting the ratio of \ion{C}{IV}/\ion{C}{III}] as possible evidence of AGN activity.

\begin{figure}
\begin{center}
\includegraphics[width=0.8\columnwidth]{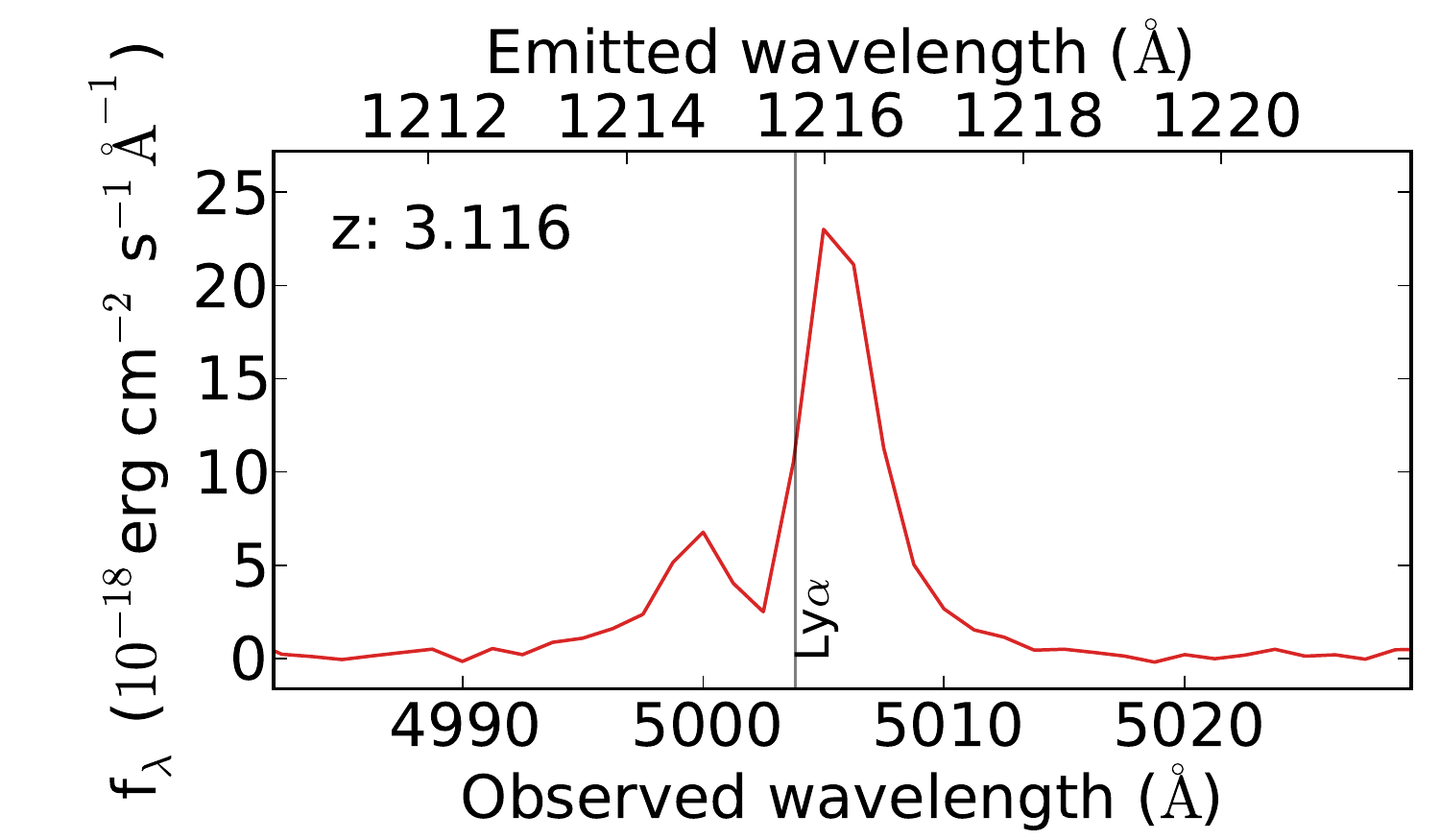}
\includegraphics[width=0.8\columnwidth]{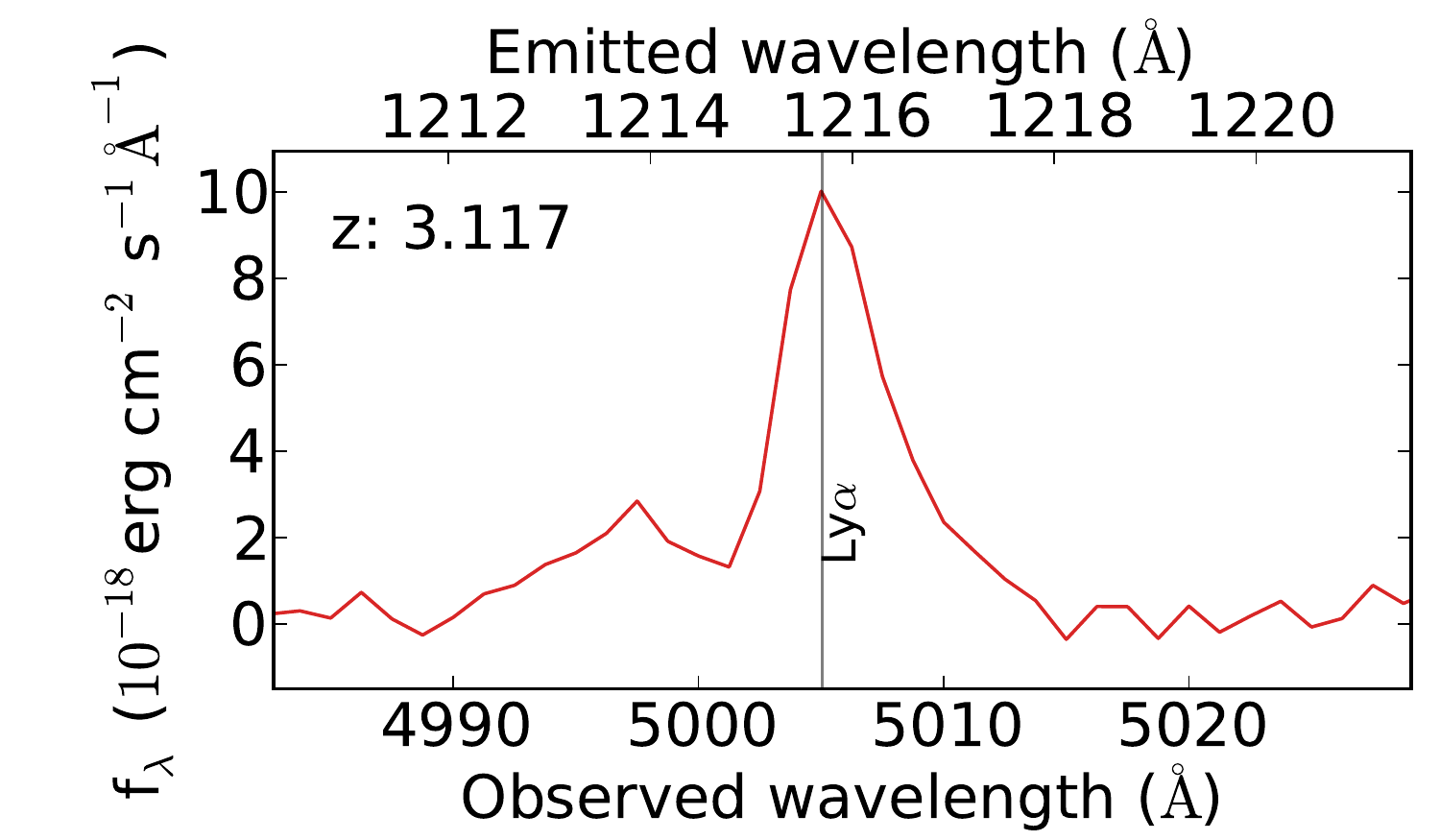}
\includegraphics[width=0.8\columnwidth]{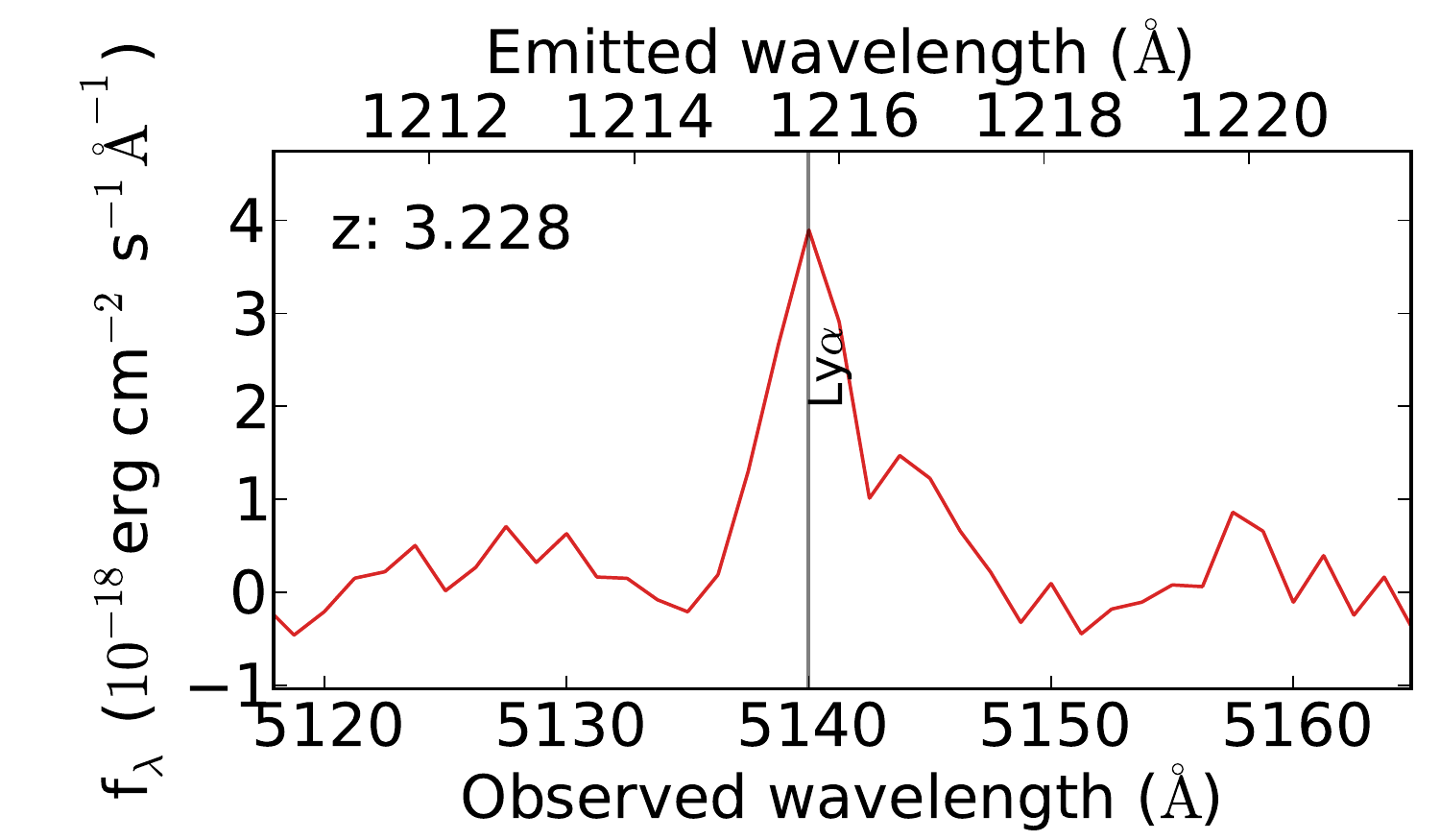}
\includegraphics[width=0.8\columnwidth]{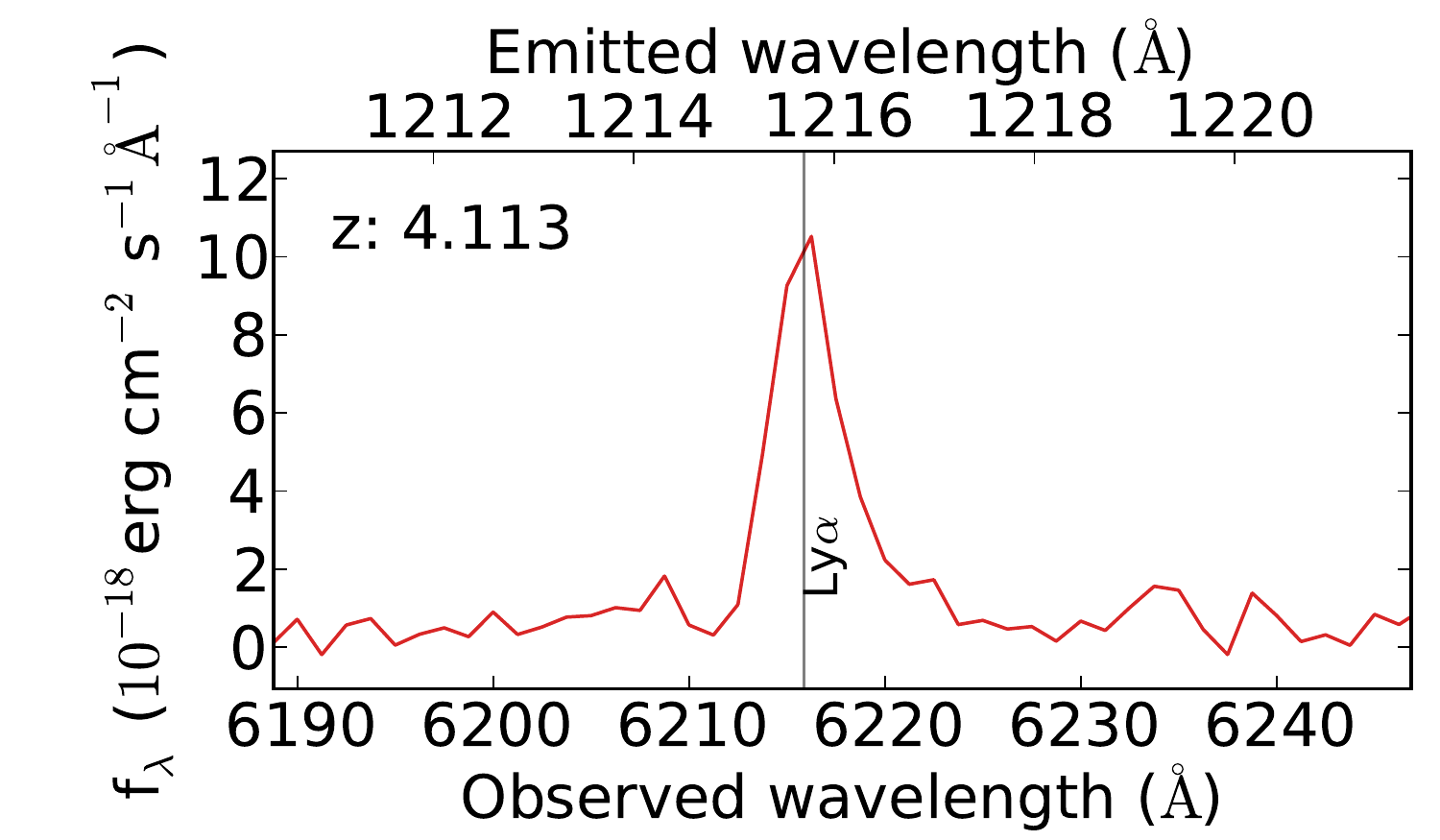}
\includegraphics[width=0.8\columnwidth]{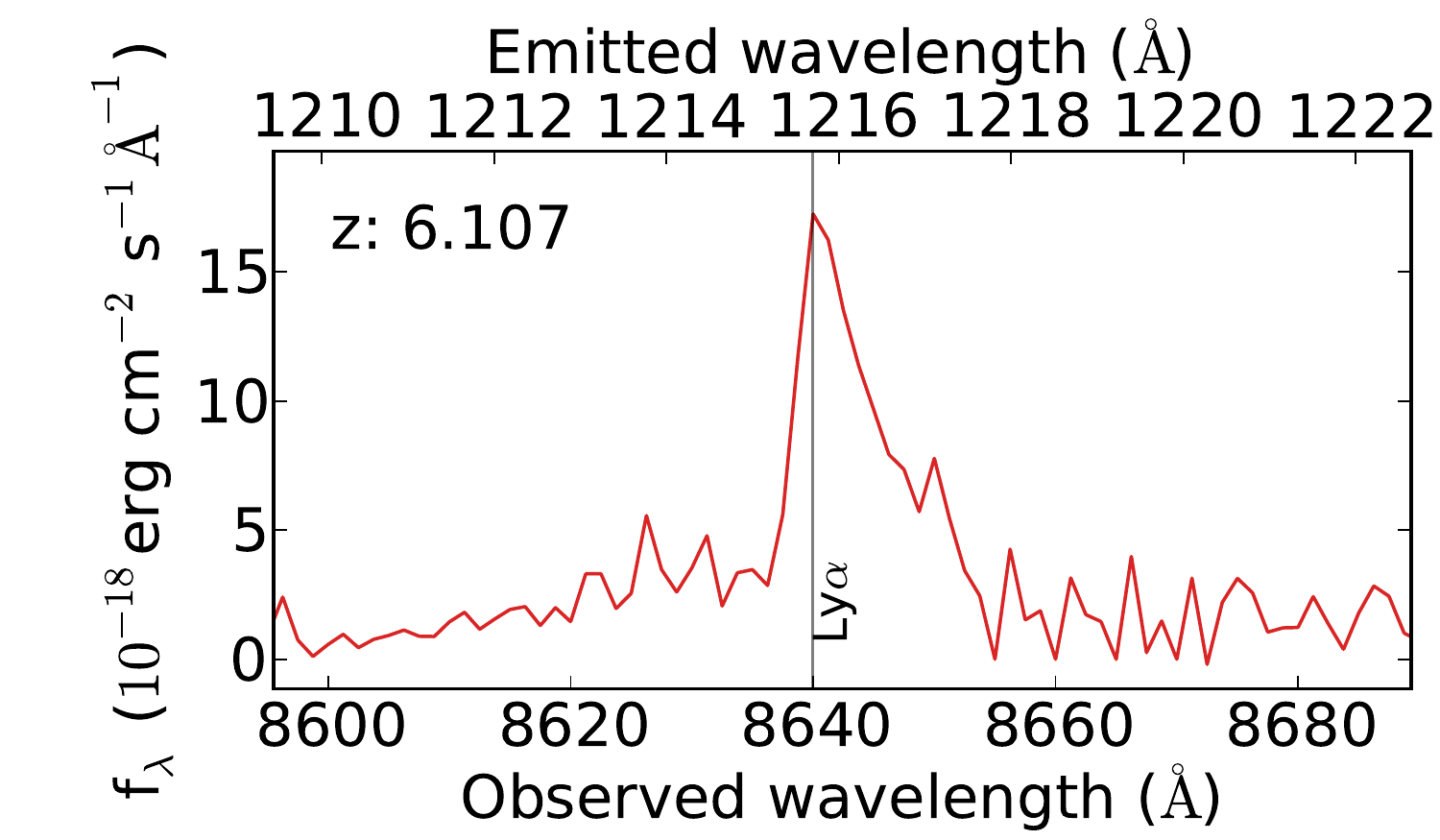}
\caption{Full-resolution MUSE 1D spectrum of sources (top to bottom)
 49, 50, 51, 52, and 53, with wavelength on the horizontal axis and the flux on the
vertical axis. The spectra are zoomed in on the \Lya line, which is indicated by the vertical black line. 
For source 49, we determined the redshift based on other UV-emission lines, while for the other
sources we used \Lya to determine $z$.
For sources 49, 52, and 53 the different images of the
same source are stacked to optimise the signal-to-noise ratio. The fluctuations and non-zero continuum 
around \Lya in object 53 are the result of locally increased noise. \label{fig:Lya}} 
\end{center}
\end{figure}

By inspecting the profile of \Lya emission for this galaxy, we can derive some important properties.
First, the profile is clearly double peaked and asymmetric. Second,
the peak \Lya emission is at 1216.15 \AA, see Fig. \ref{fig:Lya}, resulting in a redshift of 100 \kms compared
to the other UV-emission lines. This velocity difference is significantly smaller than 
observed in $z=2-3$ Lyman break galaxies \citep[$\sim400 \kms$; e.g.][]{Shapley2003,Steidel2010},
but comparable to the differences found for $z=5.7$ \Lya Emitters (LAEs) (Diaz et al., in prep.) and LAEs
in several other studies \citep[e.g.][]{Christensen2012a,Hashimoto2013,Shibuya2014}.
Third, we measure that the velocity difference between the red
and the blue peak is $\Delta v$ = 350 km s$^{-1}$, half of what 
is found in star forming \citep{Kulas2012} and massive galaxies 
\citep{Karman2014} at $z\sim3$. The center of the absorption trough is slightly blueshifted 
compared to the observed emission lines. Last, we see that the spatial profile
of the \Lya peaks is very homogeneous with a constant velocity over both sources
and two clear elliptical shapes. However, this homogeneous shape is likely the 
result of the seeing during the observations.

We also identified three additional galaxies based on
\Lya emission that were not reported before, and we show the stacked full-resolution 1D spectra of
each of these galaxies in Fig.~\ref{fig:Lya}. The first galaxy, ID 51, lies at
 a short distance from source~49 at 5140 \AA. The line has a clear asymmetry with
 a red tail and no other features at other wavelength are evident. Therefore, we determine 
the redshift to be $z=3.228$. 

The second source, ID 50, is a line emitter with 
an arc-like structure, see Fig. \ref{fig:fov_muse}d. This emission line has a central 
wavelength of 5006.3 \AA\ and no other emission is found for this 
source. Assigning \Lya to this emission results in a redshift $z=3.117$. There is a clear
red and blue peak in this source, and we measured a velocity difference of $500 \kms$ between the two peaks.

The third new high redshift galaxy that we found is source 52. It is located slightly below the arc identified
as 7a by \citet{Monna2014} and has two slightly elongated 
emission features at 6216.3 \AA, see Fig. \ref{fig:fov_muse}b. The profile of this emission line 
is asymmetric with a red wing, but we do not find a second peak, see Fig. \ref{fig:Lya}.
If this emission corresponded to a restframe 
optical nebular emission line, e.g. \ion{H}{$\alpha$}, other emission lines should be visible in 
our data. Because we do not find any other emission line and given the profile of the emission,
we conclude that these emission lines must be \Lya and that both detections are very likely
new multiply lensed images of the same galaxy at $z=4.113$. The location of the images are
consistent with submitted lens models for the Frontier Fields, and a fainter third image is predicted
north-west of the field. 

The highest redshift source that we found is a galaxy at redshift
$z=6.107$. This galaxy is discussed in detail as a quintiply lensed galaxy by 
\citet{Balestra2013}, \citet{Richard2014a}, and \citet{Johnson2014}. 
We identified both lensed images (53a\&b) that were reported before
in our field, and show the stacked \Lya profile in Fig. \ref{fig:Lya}.
In Fig. \ref{fig:fov_muse}a we have marked a tentative additional source that has the same
redshift as the multiply lensed $z=6.107$ galaxy. Although the signal at this location 
is weak, the source appears and disappears at identical wavelength (around 8641.3 \AA)
as the sources 53a\&b. This additional image is not predicted by the lens models,
and there is no clear detection in any of the {\em HST} images. If this is the same source,
we can use the ratio of the \Lya line flux in different images to calculate the magnitude at different
 wavelengths. We find a flux ratio of 5-10, and when we compare the expected magnitudes to the limits 
in the {\em HST} images, we find that for the lower ratio the source should be visible in the $F105W$ band,
 and for the higher ratio we expect a 2$\sigma$ measurement. We caution that this is a tentative detection, 
and we call this galaxy 53c, but we emphasise that it still
needs to be confirmed whether this is indeed another image of the multiply lensed source, or a companion galaxy.

The small 
difference in redshift that we and \citet{Richard2014a} find compared to \citet{Balestra2013}, 
$z=6.107$ versus $z=6.110$, is likely due to a 
difference in spectral resolution and the proximity of a weak sky line. Contrary to 
\citet{Balestra2013} we do not detect any continuum in any of the 
images. Although the exposure time is comparable, the brightest
image is close to the edge of our field, and this causes an increased 
noise level in our datacube. Therefore, the weak continuum that was 
observed previously, is lost to the locally increased noise in our observations.

\section{Narrow-band images}
\label{sec:narrowband}

We exploited the large wavelength range of MUSE to construct
narrowband images of $\sim5-10$~\AA\  wide from 6000~\AA\ to 
9300~\AA. We created these in order to look for emission
lines or weak continuum detections corresponding 
to additional high redshift galaxies, which were not identified through
visual inspection of the datacube. We 
placed the narrowbands around the atmospheric lines, excluding 
wavelengths with bright atmospheric emission, and 
determined their width to optimise the number of bands.
Because the width of these bands is comparable to the width
of \Lya emission lines, we expect that high redshift
LAEs only show up in one to three bands.

We used the photometric catalogues created by the CLASH Image Pipeline to 
compile a list of additional high redshift candidates in our field. 
We ran {\sc sextractor} on the 193 MUSE narrow-band images,
and found no detections that lie within 1\arcsec\ to the positions
of these high redshift candidates.
We found five new galaxies that occur in many of the redder images using
this approach, that are not close to the location of any high-redshift candidate. 
Because they show continuum at longer wavelengths without detections
at shorter wavelengths, it is likely that the 4000 \AA\
or Lyman break is observed. The 4000 \AA\ break only falls inside of the MUSE
wavelength range at low redshift ($z<1.2$), while the Lyman break is inside the
MUSE wavelength range at high redshift ($z>4.2$). Therefore, depending
on which break is observed, these galaxies could be either at high or low redshift. 
We labelled these sources in
Table \ref{tab:results2} with IDs higher than 60.


\section{Summary and conclusions}
\label{sec:discussion}

We presented VLT/MUSE observations of the cluster AS1063, and
provided redshift measurements for 53 galaxies. We measured redshifts 
for 34 cluster galaxies, including 29 new determinations, and 17
galaxies at higher redshifts. Five of the detected galaxies
show two or three lensed images in our field, and they are often seen
on arcs. 

We found that almost all cluster galaxies show spectra that are
representative of old stellar populations, with a strong
Balmer break and prominent \ion{Ca}{II} absorption lines. Only three out of the
34 show emission lines that indicate star formation activity. 

At intermediate redshifts ($0.4<z<1.5$), we identified 11 galaxies. All of these
galaxies show clear [\ion{O}{II}] emission lines and additional
emission lines if they are within the MUSE wavelength range, but
only two show a clear continuum. As [\ion{O}{II}] moves outside 
of the spectral range at a redshift of $z>1.5$ and \Lya only
falls inside at redshifts $z>2.9$, we do not find any galaxies
in between these redshifts. Recently, \citet{Stark2014} 
argued that \ion{C}{III}]~$\lambda 1909$ \AA\ is the best alternative in the UV
to detect low-metallicity low-mass star-forming galaxies
if \Lya is unavailable \citep[see also][]{Balestra2010}. The spectral resolution of MUSE is 
sufficient to resolve the \ion{C}{III}] doublet, disentangling
among different emission lines, even with single line detections,
as demonstrated by \citet{Richard2014b}.
As \ion{C}{III}] falls inside the 
wavelength coverage of MUSE at redshifts $1.5<z<3.9$,
deeper observations can fill the currently observed redshift gap at
$1.5<z<2.9$ with \ion{C}{III}] emitters. 
We used a LAE at $z=2.577$ (source 48, see Balestra et al., in prep) to confirm this,
and find an emission line at the wavelength corresponding to
\ion{C}{III}] $\lambda 1907$. 

At $z>3$ we detect five galaxies, of which three were not
reported before. This includes the detection of a new multiply 
lensed galaxy at $z=4.113$, with coordinates consistent
with lens model predictions. We identified two of the images reported 
by \citet{Balestra2013} as a quintiply lensed galaxy at
$z=6.107$. 

It is interesting to note that \citet{Boone2013} reported an excess
of 870~$\mu$m emission in this cluster. While two heavily star-forming
galaxies (objects 6 and 38) can explain the infrared emission from
3.6 -- 500~$\mu$m, and most of the south-west 870 $\mu$m emission,
they cannot explain the large 870~$\mu$m flux in the north-east. \citet{Boone2013}
argued that this excess could be explained by either substructures in the 
 Sunyaev-Zel’dovich effect, or a sub-mm galaxy at $z>4$. In particular,
they suggest that the multiply lensed $z=6.107$ galaxy or a companion
galaxy could be the optical counterpart to this excess. 

In our field we found three possible counterparts to such $z>4$
sub-mm galaxy. First, the quintiple-lensed $z=6.107$ galaxy (sources 53a\&b) 
was shown to have a very blue UV-slope
by \citet{Balestra2013}, which they argued is incompatible with the amount of
dust needed for a sub-mm galaxy. The second possible counterpart is object 53c,
as this could be a companion galaxy to the quintiply-lensed $z=6.107$ galaxy. 
However, the lensing models do not predict any
image of this source to the NE where most of the sub-mm excess is found. 
Therefore, we find it unlikely that this galaxy is a major contributor
to the 870~$\mu$m flux. The last possible $z>4$ contributor to this flux
is the newly detected $z=4.113$ double-lensed LAE. For this galaxy, 
the lens models do not predict an image at the NE either, making this source
equally unlikely to be significantly contributing to the 870~$\mu$m flux.
Therefore, we believe that it is unlikely that any of the galaxies 
discussed in this paper are the optical counterpart to the sub-mm source 
detected by \citet{Boone2013}.  

For one lensed \Lya emitting galaxy at $z=3.116$,
we find additional emission lines that correspond
to \ion{C}{IV}, \ion{He}{II}, \ion{O}{III}], and \ion{C}{III}],
with flux ratios indicative of an AGN. We remark that although
we favour the AGN interpretation, it is possible that these emission
lines are coming from a cluster of hot {\em O}-stars.
\citet{Ouchi2008} studied the AGN fraction as a function of \Lya
luminosity for LAEs at $z=3-4$. We determined the \Lya
flux of this object, and compare to the results of \citet{Ouchi2008}
to assess how common it is.
We measured a \Lya flux of $f=1.4\times 10^{-16}$ erg s$^{-1}$ cm$^{-2}$,
corresponding to an observed luminosity of 
$L\approx5\times 10^{43}$ erg s$^{-1}$ at this redshift. From the magnification maps 
distributed by the {\em HST} Frontier Fields programme, we
obtained a magnification factor varying from 5 to 20, depending
on the lensing model. This results in an intrinsic luminosity of
$L=2.5-10\times 10^{43}$ erg s$^{-1}$.
\citet{Ouchi2008} found that at least 10 per cent of LAEs with 
similar \Lya luminosities host an AGN.

All of the high-$z$ galaxies show a clearly
asymmetric \Lya line profile, with a red wing which is more extended
than the blue side, and two show a large red peak in combination with 
a smaller blue peak. The velocity difference between the two peaks
is significantly smaller than seen in most previous studies. \citet{Christensen2012b}
find a similar peak velocity difference in their lensed sources. 
A possible explanation is that by observing lensed sources, we find less
massive galaxies with smaller gas-expansion speeds, and as a result
a less blueshifted second peak.
The \Lya emission lines all show a regular spatial
elliptical shape, which are aligned with the detected arcs. As these
shapes could be the result of the lensing and seeing conditions,
we can unfortunately not draw any conclusions about the intrinsic shape. 
With the future implementation of Adaptive Optics at MUSE,
these structures should become clearly resolved.

\begin{figure}
\begin{center}
\includegraphics[width=\columnwidth]{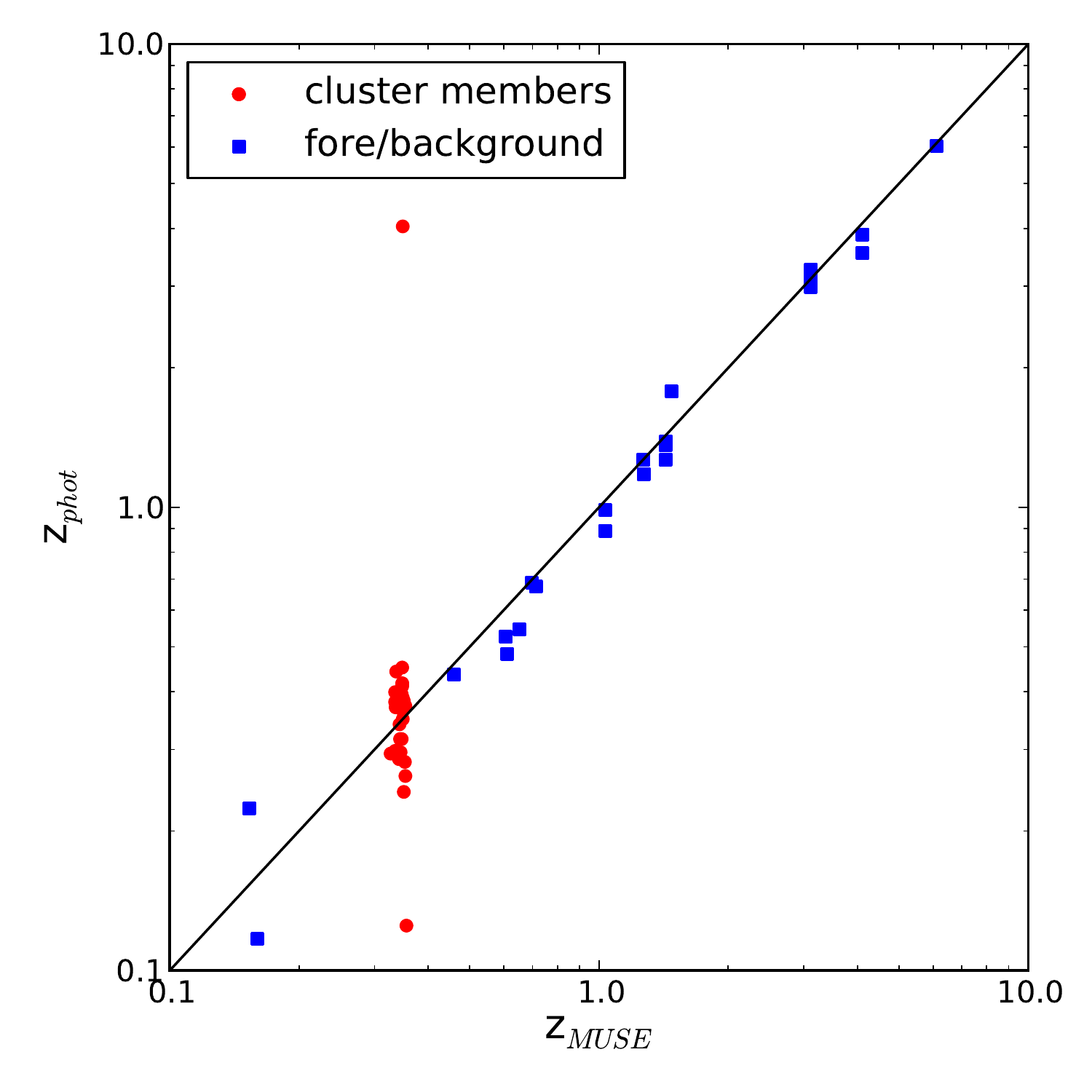}
\caption{Comparison of the photometric redshifts as calculated by the CLASH team based on {\em HST} 
multiband photometry, with the MUSE spectroscopic redshifts measured in this study. Red circles correspond to
cluster members, while the blue squares correspond to foreground and background galaxies. Objects 43b and
51 are not deblended in the {\em HST} images, objects 19 and 53a are missing several bands, and object 53c
has no {\em HST} detections, and are therefore all excluded from this plot.
\label{fig:zpzs}}
\end{center}
\end{figure}

It is interesting to compare the photometric redshifts, as estimated by the CLASH
team, with the MUSE spectroscopic redshifts. We show this comparison in Fig. \ref{fig:zpzs}.
The agreement found for the background galaxies is very good, the cluster members
instead show a larger degree of scatter. 

In this study, we show the breadth of science that a single MUSE pointing, with
modest exposure time, is able to address. At high-$z$, it allows for the detection and 
characterisation of galaxies through \Lya detection, rest-frame UV lines, and morphology. At low and
intermediate redshifts,
it enables the construction of rotation maps, the study of internal dynamics, the determination of the
ionisation source in the galaxy through emission line ratios, and accurate stellar population studies. 
At low redshifts, the properties will be obtained in high detail, and in addition, 
star formation rates can be determined from \ion{H}{$\alpha$}.

This Article shows the strength of using MUSE in both blind and targeted searches for high-redshift
galaxies in magnifying clusters. The additional advantage of MUSE, when observing magnifying clusters, 
is that one simultaneously observes the whole Einstein volume of the cluster in one pointing.
MUSE provides both a high spatial and spectral 
resolution, has a large FOV and spectral range. These properties make it ideal to study line profiles and look for 
additional features without the problem of source confusion. As the performance
is good over long times, it is well suited for long exposures in empty sky fields.
MUSE will therefore provide a very large number of galaxies with redshift identifications
up to a redshift of $z<6.6$, and is expected to significantly increase our understanding
of the formation and evolution of galaxies at $z>3$.

\begin{acknowledgements}
 
Based on observations made with the European Southern Observatory 
Very Large Telescope (ESO/VLT) at Cerro Paranal, under programme ID 60.A-9345(A).
This work utilises gravitational lensing models produced by PIs Brada\'c, Ebeling, 
Merten \& Zitrin, Sharon, and Williams funded as part of the HST Frontier Fields
 programme conducted by STScI. STScI is operated by the Association of 
Universities for Research in Astronomy, Inc. under NASA contract NAS 5-26555.
The lens models were obtained from the Mikulski Archive for Space Telescopes (MAST).

The authors thank the anonymous referee
for his/her helpful comments that improved the clarity of this article.
The authors also thank Thomas Martinsson for help with the MUSE pipeline,
Benjamin Clement for support in designing the observations, and 
Elena Valenti and Vicenzo Mainieri for technical support during the observation
planning and data reduction phase. 

The Dark Cosmology Centre is funded by the DNRF.

\end{acknowledgements}

\bibliographystyle{aa}
\bibliography{muse.bib}

\end{document}